\documentclass[journal]{IEEEtran}

\usepackage{pstricks}
\usepackage{mathptmx}
\usepackage{cite}
\usepackage[dvips]{graphicx}
\usepackage[cmex10]{amsmath}
\usepackage{algorithm}
\usepackage{algpseudocode}

\usepackage{caption}
\usepackage{subcaption}

\usepackage{multirow}
\usepackage{url}
\usepackage{color}
\usepackage{threeparttable}
\usepackage{array}
\usepackage[top=0.8in,bottom=0.8in,left=0.6in,right=0.6in]{geometry}
\usepackage[colorlinks, citecolor=blue,linkcolor=blue,urlcolor=blue]{hyperref}                     %
\graphicspath{{./figs/}}

\begin{document}
\newtheorem{property}{\textbf{Property}}
\newtheorem{definition}{\textbf{Definition}}
\newtheorem{lemma}{\textit{Lemma}}
\newtheorem{theorem}{\textbf{Theorem}}
\newtheorem{corollary}{\textbf{Corollary}}
\newcommand{\revise}[1]{\textcolor{black}{#1}}
\newcommand{\minrev}[1]{\textcolor{black}{#1}}

\title{Generalized Fault-Tolerance Topology Generation for Application Specific Network-on-Chips}

\author{
Song Chen,~\IEEEmembership{Member,~IEEE},
\revise{Mengke Ge,}
Zhigang Li,
Jinglei Huang,
Qi Xu,
and \revise{Feng Wu,~\IEEEmembership{Fellow,~IEEE}}
\thanks{This work was partially supported by the National Natural Science Foundation of China (NSFC) under grant Nos. 61874102 and 61732020, \minrev{Beijng Municipal Science \& Technology Program under Grant Z181100008918013,} and the Fundamental Research Funds for the Central Universities under grant No. WK2100000005.
The authors would like to thank Information Science Laboratory Center of USTC for the hardware \& software services.
}
\thanks{
S. Chen, M. Ge, Z. Li, and F. Wu are with the School of Microelectronics, University of Science and Technology of China (USTC), China; (Email: songch@ustc.edu.cn). S. Chen and F. Wu are also with USTC Beijing Research Institute, Beijing, China

\revise{J.~Huang is with State Key Laboratory of Air Traffic Management System and Technology, China (email:huangjl@mail.ustc.edu.cn).}

\revise{Q.~Xu is with the School of Electronic Science and Applied Physics, Hefei University of Technology, China (email: xuqi@hfut.edu.cn).}
}
}

\markboth{IEEE TRANSACTIONS ON COMPUTER-AIDED DESIGN OF INTEGRATED CIRCUITS AND SYSTEMS, VOL. , NO.}
{S. Chen \MakeLowercase{\textit{et al.}}: \textsc{Fault-Tolerance Application-Specific NoC Topology Synthesis Generalized Fault-Tolerance Topology Generation for Application Specific Network-on-Chips}}

\maketitle
\begin{abstract}
The Network-on-Chips based communication architecture is \revise{a promising candidate} for addressing communication bottlenecks in \minrev{many-core} processors and neural network processors.
In this work, we consider the generalized fault-tolerance topology generation problem, where the link (physical channel) or switch failures \revise{can happen}, for application-specific network-on-chips (ASNoC).
With a user-defined maximum number of faults, $K$, we propose an integer linear programming (ILP) based method to generate ASNoC topologies, which can tolerate at most $K$ faults in switches or links.
Given the communication requirements between cores and \revise{their floorplan}, we first propose a convex-cost flow based method to solve a core mapping problem for building connections between \revise{the cores and switches}.
Second, an \revise{ILP} based method is proposed to solve the routing path allocation problem, where $K+1$ switch-disjoint routing paths are allocated for every communication flow between \revise{the} cores.
Finally, to reduce switch sizes, we propose \revise{sharing} the switch ports for the connections between the cores and switches and formulate the port sharing problem as a clique-partitioning problem, which is solved by iteratively finding the maximum cliques.
Additionally, we propose an \revise{ILP}-based method to \revise{simultaneously solve} the core mapping and routing path allocation problems when only physical link failures are considered.
Experimental results show that the power consumptions of fault-tolerance topologies increase almost linearly with $K$ because of the routing path redundancy for fault tolerance. When both switch faults and link faults are considered, port sharing can reduce the average power consumption of fault-tolerance topologies with $K=1$, $K=2$ and $K=3$ by \minrev{18.08}\%, \minrev{28.88}\%, and \minrev{34.20}\%, respectively.
When considering only \revise{the} physical link faults, the experimental results show that compared to the FTTG \revise{(fault-tolerant topology generation)} algorithm, the proposed method reduces power consumption and hop count by 10.58\% and  6.25\%, respectively; compared to the DBG \revise{(de Bruijn Digraph)} based method, the proposed method reduces power consumption and hop count by 21.72\% and 9.35\%, respectively.
\end{abstract}



\section{Introduction}
\label{sec:intro}
With the \revise{constant} scaling of semiconductor manufacturing technologies, hundreds to thousands of processing cores can be easily integrated on a single chip \cite{dac2007-manycore}.
Network-on-Chips (NoCs) have emerged as an attractive solution to the interconnection challenges of heterogeneous System-on-Chip designs \cite{intro2} \cite{acs-2006-survey} \cite{tcad-2009-survey} and neuromorphic computing systems \cite{truenorth2015} \cite{neurogrid2014} \cite{neu-noc2018} because NoCs have good scalability and enable efficient and flexible utilization of communication resources \revise{when} compared \revise{to} the traditional point-to-point links and buses. NoCs convey messages (in packets) through a distributed system of routers (sometime called switches in ASNoCs) interconnected by links, and these routers may include network interfaces for connecting cores to routers. In this work, we focus on ASNoCs\minrev{, where the customized irregular network topologies are used} because of \revise{their} low energy consumption and low area overhead \cite{intro5,ASNoC:1}.

With successive technology node shrinking, the \minrev{transistor size on chips has been scaled down to a few nanometers}, where radiation, electromagnetic interference, electrostatic discharge, aging, process variation and dynamic temperature variation are the major causes of failures in MOSFET based circuits\cite{micro2003} \cite{micro2005} \cite{spectrum2011}.
It is extremely difficult for a heterogeneous system to \revise{guarantee} long-term product reliability because of a combination of these factors.
To maintain network connectivity and correct packet-switching operations, we consider fault-tolerance issues of \revise{the} network components in ASNoCs.
NoC with regular topologies can achieve fault tolerance by providing alternative routing paths when messages or packets encounter faulty network components. However, in ASNoCs, the path diversity is greatly reduced for \revise{lowering} the energy and area overhead of the network components. Consequently, we have to introduce structural redundancies, such as switches, ports, links, and network interface, to \revise{address these} faults \cite{intro:redun1,intro:redun2,iscas2013a,iscas2014a}. Then, alternative routing paths are used for the packet switching between the cores, thus bypassing the faulty region\cite{intro:routingAlg}. Note that, generally, fault control in NoC involves two phases: fault diagnosis and \minrev{fault tolerance} \cite{toc2016}.
Our research mainly focuses on \revise{the} fault tolerance.

There are many previous works addressing the synthesis of ASNoC topologies \cite{ASNoC:1,ASNoC:2,ASNoC:3,ASNoC:4,ASNoC:5,ASNoC:6,ASNoC:7,ASNoC:8,ASNoC:9,ASNoC:10,ASNoC:11,ASNoC:12,ASNoC:13}.
However, these works \revise{rarely} consider fault tolerance in the NoC topologies. \revise{In particular}, the ASNoCs have low path diversities and \revise{cannot} work normally if any hardware faults occur \revise{in the} switches or links.
In \cite{intro6}, a fault tolerant NoC architecture was proposed\revise{, where the} cores \revise{were} linked to two switches instead of one, and a dynamically reconfigured routing algorithm was used to \revise{bypass} faulty switches.
Chatter et al.\cite{spareRouter} proposed a fault-tolerant method based on router \revise{redundancy}. They \revise{allocated} a spare router for each router for fault tolerance, which increased \revise{the consumed} power and area.
In \cite{quadSpareRouter}, the authors place\revise{d} a spare router for each $2\times 2$ router block in the mesh topology and used multiplexers to switch the faulty routers to the intact routers, which \revise{could thereby} decrease the power and area overheads compared to \cite{spareRouter}. However, this method cannot be applied to \revise{ASNoC} designs.

Tosun et al. \cite{FT} proposed a fault-tolerant topology generation (FTTG) method for ASNoC, which focused on permanent link and switch port failures.
The authors \revise{attempted} to add a minimum number of extra switches and links and use the min-cut algorithm to ensure that each switch and link \revise{were} on a cycle, which provided at least two alternative routing paths to achieve \revise{adequate} fault-tolerance.
The NoC topologies are generated in two phases. In the first phase, the links between the switches (switch topologies) are constructed, and in the second phase, the links from the cores to switches are built (core mapping). However, the switch topology and the core mapping strongly depend on each other; consequently, it is challenging for the FTTG method to effectively explore the design space of the network topologies.
Additionally, the FTTG can only generate one-fault tolerant topologies and cannot be applied \revise{toward} generating multiple-fault tolerant network topologies.

Motivated by \revise{these} arguments, we propose a method for generating ASNoC topologies \revise{with consideration of} both switch faults and physical link faults, when given the communication requirements between the cores, floorplan of the cores, and maximum number of tolerable faults, $K$. The main contributions of this work are as follows.
\begin{enumerate}
\item We propose a generalized fault-tolerant topology generation method with consideration of both switch faults and link faults. A convex-cost flow based method is used to solve \revise{the} core mapping problem \revise{for building} connections between \revise{the cores and switches, and}
an ILP based method is proposed to allocate $K+1$ switch-disjoint routing paths for each communication flow.
\item To reduce the switch sizes, we propose sharing the switch ports for the connections between the cores and switches, and prove the conditions for port sharing \minrev{on a switch}. The port sharing problem \minrev{on a switch} is formulated as a clique-partitioning problem and heuristically solved by iteratively \minrev{finding a set of maximum cliques} and solving a maximum cardinality matching problem. \minrev{Moreover, we propose a heuristic method, where a series of maximum independent set problems are solved for removing the conflicts caused by port sharing on multiple switches.}
\item Additionally, we also propose an ILP-based method to simultaneously solve the core mapping and routing path allocation problems when only the physical link failures are considered.
\end{enumerate}

Experimental results show that the power consumptions of fault-tolerance topologies increase almost linearly with $K$ because of the routing path redundancies \minrev{(See Fig.\ref{fig:PowerAnalysisWithNFT})}.
When both switch faults and link faults are considered, port sharing can respectively reduce the average power consumptions of the fault-tolerance topologies with $K=1$, $K=2$ and $K=3$ by \minrev{18.08}\%, \minrev{28.88}\%, and \minrev{34.20}\%.
When considering only the physical link faults, the experimental results show that, compared to the FTTG, the proposed method reduces power consumption and hop count by 10.58\% and  6.25\%, respectively; compared to the DBG based method, the proposed method reduce power consumption and hop count by 21.72\% and 9.35\%, respectively.

The remainder of this paper is organized as follows. Section \ref{sec:problem} formulates the $K$-fault-tolerant ASNoC topology generation problem. The overview of the proposed framework is shown in Section \ref{sec:overview}. The generalized $K$-fault-tolerant topology generation methodology is discussed in section \ref{sec:map}, \ref{sec:pa} and \ref{sec:ps}. Section \ref{sec:linkonly} discusses the generation method for link-fault-tolerance topologies. The experimental results are provided in Section \ref{sec:experiment}, followed by \revise{the} conclusions in Section \ref{sec:conclusion}.

\section{Preliminaries \& Problem Formulation}
\label{sec:problem}
\subsection{NoC Architecture}
In this work, the ASNoC architectures are assumed to support packet-switched communications with source routing and wormhole flow control \cite{noc-stanf}.
In the application-specific design, the communication characteristics are known \revise{a priori}, and hence, \minrev{a deterministic routing} strategy is used; that is, the routing path for the communications is preallocated, which accordingly determines the topology of the NoC.
The ASNoC topology architecture consists of two main components: switches and customized electrical links.
The switches are used to route packets from the source to the destination, and the routing information \revise{is included} in the packet to specify the address of the output port, to which the packet should be forwarded.
Given the communication characteristics of an application, this work focuses on the generation of network topologies by preallocating the routing paths for the communication flows.

\subsection{Problem \revise{Definition}}
\label{sec:pf}
\begin{figure}[htbp]
\small \centering
  \includegraphics[width=8.00cm]{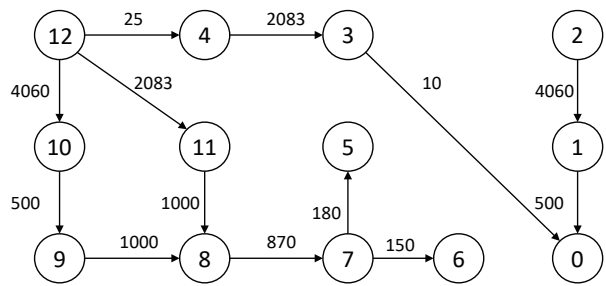}\\
  \caption{$G_{cc}$ of MP3EncMP3Dec encoder application.}
  \label{fig:mp3CCG}
\end{figure}

Let $\revise{V_{c}} = \{c_i| 1\le i \le n_{core}\}$ be the set of cores in an application. The communication requirements (or communication flow in this work) between the cores can be represented \revise{as} a directed graph, $G_{cc}$, and defined as follows.
\begin{definition}
\label{def:g_cc}
$G_{cc} = (V_{c}, E_{cc})$ is directed. An edge $(c_{i},c_{j})\in E_{cc}$ represents the communication from  $c_{i}$ to $c_{j}$. Besides, the bandwidth requirement of the communication flow from $c_i$ to $c_j$ is given by $w_{i,j}$.
\end{definition}
Fig. \ref{fig:mp3CCG} shows an example of $G_{cc}$.

In ASNoCs, the switches will be shared among the cores for data communications.
If only link failures between the switches are considered \cite{FT}, the mapping from the cores to the switches is a many-to-one relationship, which is the same as the clustering problem in the traditional ASNoC synthesis \cite{ASNoC:9}\cite{ASNoC:3}.
However, the mapping from cores to switches is a many-to-many relationship if both switch failures and link failures are considered.
Let $V_s = \{s_i| 1\le i \le n_{sw}\}$ be the set of switches.
We \minrev{use the Cartesian products $V_c \times V_s$, $V_s \times V_c$, and $V_s \times V_s$ to represent all possible connections from the cores to the switches, from the switches to the cores, and from switches to switches, respectively}.


The problem of generalized fault-tolerance topology generation for ASNoCs is defined as follows.

\noindent\textbf{Problem \minrev{statement}.}

Given a core communication graph $G_{cc}$, number of switches $n_{sw}$, floorplan of the cores, and number of tolerable faults $K$, we attempt to determine the placement of the switches and construct a $K$-fault-tolerant ASNoC topology \revise{with minimization of the power consumption of the ASNoC under the following constraints:}
\begin{itemize}
  \item the \textit{latency}  constraint \revise{$l_{i,j}$} (number of hops) for each communication flow $(c_i, c_j)\in E_{cc}$,
  \item the \textit{switch size} constraint $max\_size$, which is the maximum number of ports that a switch could support given the NoC operating frequency,
  \item and the \textit{bandwidth} constraint $BW_{max}$ for the physical links, which is the product of the NoC frequency and bit-width of the physical links.
  \end{itemize}

The ASNoC topology can be represented as a direct graph $G_{NT}(V_{NT}, E_{NT})$, where $V_{NT}= V_{c}\cup V_s$, and $E_{NT}$ includes two types of edges:
\begin{itemize}
  \item a subset of edges \minrev {$L_{cs} \subseteq V_c \times V_s \cup V_s \times V_c$}  determined by solving the \textit{core mapping} problem, corresponding to the connections between the cores and switches,
  \item and \revise{a subset of} edges $L_{ss} \subseteq \minrev{V_s \times V_s}$ determined by solving the \textit{routing path allocation} problem,  corresponding to the physical links between the switches.
\end{itemize}

In $G_{NT}$, \revise{there are $K+1$ switch-disjoint paths for each communication flow $(c_i, c_j)$ in $G_{cc}$} when both switch failures and link failures are considered.

In the $K$-fault tolerance structures, $K$ times more switch ports are connected to each core for introducing routing path redundancies.
These switch ports greatly increase the area and power consumption of switches. To reduce the number of the switch ports, we also solve a \textit{port sharing} problem for each switch, which will be discussed in details in Section \ref{sec:ps}.

\section{Overview of the Proposed Framework}
\label{sec:overview}

Given the floorplan of $n_{core}$ cores, their communication requirements represented by $G_{cc}$, and the number of switches $N_{sw}$, the placement of the switches is determined using the method in \cite{zhong2013floorplanning:11}.

As discussed in Section \ref{sec:pf}, the NoC topology generation problem mainly includes two subproblems: core mapping (CM) and routing path allocation (PA).
We first map the cores to the switches using a min-cost-max-flow algorithm in Section \ref{sec:map}. Second, the routing path allocation is solved using an ILP-based method in Section \ref{sec:pa}.
If we fail \revise{to find} $K+1$ switch-disjoint paths for all the communication flows under the given constraints, the number of switches \revise{is increased} by one and \minrev{the generalized fault-tolerance topology generation} problem is solved again.
This procedure is repeated until all the communication flows \revise{have} $K+1$ switch-disjoint routing paths.

To generate \revise{the $K$-fault-tolerant topology}, we connect each core to at least $K+1$ switches \minrev{by $K+1$} ports, which greatly increases the power and area overheads of the switches.
However, for \minrev{each flow of} each core, only \minrev{one of all the $K+1$ ports} work for data communication. Consequently, the switch ports connecting different cores could be shared using multiplexers.
 In Section \ref{sec:ps}, we prove the conditions for port sharing \minrev{on a switch} and propose a clique-partitioning formulation for the problem, which is solved using a heuristic method. \minrev{Moreover, a heuristic method is proposed to remove the conflicts of routing path selection, caused by port sharing on multiple switches.}

\begin{figure}[htbp]
\small \centering
  \includegraphics[width=9.00cm,height=5.30cm]{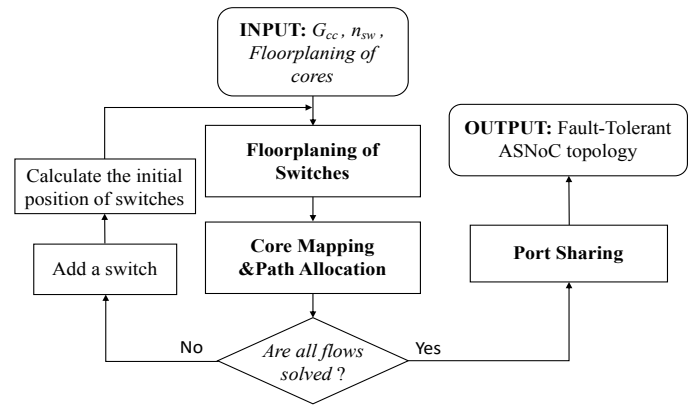}\\
  \caption{\revise{Overview} of the proposed framework.}
  \label{fig:overview}
\end{figure}

Fig.\ref{fig:overview} illustrates the overall flow of generating fault-tolerance ASNoC topologies.

\section{Fault-Tolerance Topology Generation}
\subsection{Core Mapping}
\label{sec:map}
To generate $K$-fault-tolerant topology, we connect each core to \minrev{at least} $K+1$ switches, and many switch ports are introduced for connecting the cores, accordingly.
For a switch, the area \revise{increases} quadratically with the port number, and the power \revise{increases} superlinearly with the switch size.
Consequently, a convex-cost flow based method is used to generate a core mapping with evenly distributed core--switch connections.

 \revise{In the core mapping stage}, we build connections from the source cores of the communication flows to the switches and connections from the switches to the sink cores of the communication flows.
Here, we consider building connections from the source cores to the switches.
The connections from the switches to the sink cores are built similarly.

To build a convex-cost flow model, we \minrev{construct} a directed graph \revise{$G_{cs}(V_{cs},E_{cs})$}, where $V_{cs} = V_{c} \cup V_{s} \cup \{b,t\}$ and $E_{cs} = \minrev{V_c \times V_s} \cup \{V_{s}\to t\} \cup \{b \to V_{c}\}$.


The capacity of an edge $(\revise{b}, c_i) \in \{\revise{b} \to V_{c}\}$ is set to $K+1$ if there is an outgoing communication flow from $c_i$ and is set to $0$ otherwise.
{The capacities of the edges in $\{V_{s}\to t\}$ are set to \revise{$N_{cs} = \lfloor{n_{core} * (K+1)/n_{sw}}\rfloor +1$}, which is close to the average number of input ports that is used to connect cores on a switch. All the other edges have a capacity of $1$.}



The edges in $\{\revise{b} \to V_{c}\}$ have zero cost.  we can map on one switch;
For an edge $(c_i, s_j) \in  \minrev{V_c \times V_s}$, the cost is defined as $E_{bit} \times D_{c_i, s_j} \times \sum_{k:(c_i,c_k)\in E_{cc}}{w_{i,k}}$, where $D_{c_i, s_j}$ is the distance between core $c_i$ and switch $s_j$ and $E_{bit}$ (set to 0.5 in this work) is the bit energy of unit wire length ($1mm$).

 \revise{To make the connections from cores to switches evenly distributed among the switches,} the costs of the edges in $\{V_{s}\to t\}$ are defined \revise{to be} a function of \revise{the number of} flows \revise{$x$} on the edges $c_{(s_j,t)}(x)$, which \revise{corresponds to} a piecewise linear and convex \revise{function for the flow costs}.
{Let $0=d_{0}\le d_{1}\le \cdots \le d_{N_{cs}}$ denote the breakpoints of the piecewise function and the costs vary linearly in the interval $[d_{i-1},d_{i}], 1\le i\le N_{cs}$.}
In this work, \revise{the edge cost function} $c_{s_j,t}(x)=10x$ and the interval between adjacent breakpoints is set to $1$. \revise{Consequently, the flow cost is calculated as $10x^2$.}
Such a convex-cost flow problem can be easily transformed into a traditional min-cost flow problem \cite{minCostFlowOfNetworkFlows}.

According to the solution to the convex cost flow model, the edges in $\minrev{V_c \times V_s}$ that have non-zero flow will be selected as the connections from cores to switches.


After we map the cores to switches, we have determined the connections between the cores and the switches, denoted as $L_{cs} (\subseteq \minrev{V_c \times V_s \cup V_s \times V_c}$). Hereafter, for a communication flow $(c_i, c_j)$ in $G_{cc}$, we define the switch ports connected to the source core $c_{i}$  as \textit{\textbf{core inports}} and the switch ports connected to the sink core $c_{j}$ as \textit{\textbf{core outports}}.

\minrev{An example of core mapping for $G_{cc}$ in Fig.\ref{fig:mp3CCG} is shown in Fig.\ref{fig:resultOfPA}}, where $K=1$. \minrev{Each source core is connected to multiple \textit{core inports} and multiple \textit{core outports} respectively through a demultiplexer (DEMUX) and a multiplexer (MUX).}

\subsection{ILP based Path Allocation}
\label{sec:pa}

To generate $K$-fault tolerant topologies, we have to find $K+1$ alternative switch-disjoint (\textbf{node-disjoint}) routing paths in \minrev{a complete graph of switches $G_s(V_s, V_s\times V_s)$}  for each communication flow considering the costs of switches and links.
To reduce the internally node-disjoint paths problem to an edge-disjoint paths problem \cite{schrijver}, which is easily formulated as a constrained min-cost multi-flow problem, we perform node splitting on $G_{s}$ and extend the graph for routing path allocation.
Each switch node $u\in V_{s}$ is split into two nodes $u$ and $u'$.
A directed graph $G_{\revise{pa}}(V_{\revise{pa}},E_{\revise{pa}})$ is constructed as follows.
\begin{itemize}
	\item{$V_{pa}=\revise{V_c} \cup V_{s} \cup V'_{s}$, where $V'_{s}$ is the split node set of $V_{s}$.}
	\item{$E_{pa}=\revise{L_{cs}} \cup E_{split}\cup E_{link}$, where $E_{split}=\{(u,u')|u\in V_{s} \land u'$ is the corresponding split node of $u$\} and $E_{link}=$\{$(u',v)|(u,v)\in \minrev{V_s\times V_s} \land u'$ is the corresponding split node of $u$\}. If there is a directed edge from $u$ to $v$ in $\minrev{V_s\times V_s}$, a corresponding directed edge from $u'$ to $v$ is added in $E_{pa}$.}
\end{itemize}

\revise{In the following, we discuss how to find $K + 1$ edge-disjoint routing paths in $G_{pa}$ for all the communication flows.}

\subsubsection{Computation of Switch Power and link power}
\label{subsec:power-model}
\revise{The switch power depends on the switch size, which includes the number of input ports and output ports.
To calculate the size of switches, we introduce two types of binary variables $x_{uv}^{i,j,k}$ and $d_{uv}$.
 $x_{uv}^{i,j,k} = 1$ indicates that the $(k+1)$-th routing path of the communication flow $(i,j)$ goes through the edge $(u,v)\in E_{pa}$, $0\le k \le K$; otherwise, $x_{uv}^{i,j,k} = 0$.
$d_{uv} = 1$ indicates that there is at least one communication flow going through the edge $(u,v)\in E_{pa}$, and accordingly a physical link exists between the switches $u$ and $v$; otherwise, $d_{uv} = 0$.   $d_{uv}$ is calculated based on binary variables $x_{uv}^{i,j,k}$ as follows \cite{XuQi}.}
\begin{equation}\label{eq:link:degree}
d_{uv} = min\{\sum_{(i,j)\in E_{cc}}\sum_{k=0}^{K}x^{i,j,k}_{uv}, 1\}, \forall (u,v)\in E_{\revise{link}}.
\end{equation}

Then, \revise{the size of a switch $u$, including} the input port number $ip_{u}$ and output port number $op_{u}$, can be calculated as follows.
\begin{equation}
\label{eq:s:portNum}
\begin{split}
ip_{u}=\sum_{v:(v,u)\in E_{\revise{link}}} d_{vu} + cip_{u};~
op_{u}=\sum_{v:(v,u)\in E_{\revise{link}}} d_{uv} + cop_{u}.
\end{split}
\end{equation}

$cip_{u}$ and $cop_{u}$ respectively represent the number of \textit{core inport}s and \textit{core outport}s\revise{, which have been determined in the core mapping stage}.
\revise{Consequently, the power consumption of the switches are estimated using Orion 3.0 \cite{orion3} as follows\cite{ASNoC:12}.
\begin{equation}
\label{eq:power-sw}
P_{\revise{sw}}(ip_u,op_u)= \revise{T_{sw}}[ip_u]+ C_{sw}*op_u
\end{equation}
where \revise{$T_{sw}$} is a table mapping the input port number to power consumption, and $C_{sw}$ is a constant.}

\revise{
The link power of a communication flow $(i,j)$ on an edge $(u,v)\in E_{link}$ is determined by the communication requirement $w_{i,j}$ and the physical distance between $u$ and $v$, $D_{u,v}$.
Considering the cost of opening new physical links, an extra cost $C_{pl}$ is introduced if the physical link between $u$ and $v$ does not exist.
Therefore, the link power $P_{sw}(i,j,u,v)$ is calculated as follows\cite{huang2015lagrangian}.
\begin{equation}
\label{eq:cost-1}
P_{\revise{link}} = \sum_{(u,v)\in E_{\revise{link}}}\sum_{(i,j)\in E_{cc}} (E_{bit}\cdot w_{i,j}\cdot D_{u,v}\cdot x_{uv}^{i,j,0} + d_{uv}\cdot C_{pl}),
\end{equation}
where $E_{bit}$ represents the bit energy of the electrical link.}

\subsubsection{ILP Formulation for Routing Path Allocation}
\label{subsec:ilp-sl}
To simplify the discussion, the communication flows in $G_{cc}$ are relabeled as $(i,j)=(c_{i},c_{j})\in E_{cc}$ (Definition \ref{def:g_cc}).
The routing path allocation for ASNoCs with $K$-fault tolerance can be formulated as a \revise{constrained} multiple flow problem \revise{$G_{pa}$} as follows.
\begin{subequations}
\label{eq:ilpall}
\allowdisplaybreaks[4]
\begin{align}
 Min~& \revise{\sum_{u\in V_s}P_{sw}(ip_u,op_u)} + \revise{P_{link}} \tag{\ref*{eq:ilpall}}\\
    \textrm{s.t.}
    &\sum_{v:(u,v)\in E_{\revise{pa}}}x_{uv}^{i,j,k} - \sum_{v:(v,u)\in E_{\revise{pa}}}x_{vu}^{i,j,k}= \nonumber
    \left\{
    \begin{array}{ll}
        1,      & \mbox{if}~ u=\revise{c_{i}}; \\
        0,      & \mbox{if}~ u\in \revise{V_{s}};\\
        -1.     & \mbox{if}~ u=\revise{c_{j}}; \\
    \end{array} \right.  \nonumber\\
    &~~~~~~~~~~~~~~\forall ~ k \in [0,K],\forall ~ (i,j) \in E_{cc},      \label{eq:s:unitFlow} \\
    & \sum_{k=0}^{K}x^{i,j,k}_{uu'}\leq 1, \forall (u,u') \in E_{\revise{split}},\forall ~ (i,j) \in E_{cc}, \label{eq:s:edgeDisjoint}\\
    &\sum_{(u,u')\in E_{split}}x^{i,j,\minrev{k}}_{uu'}\leq l_{i,j},\minrev{\forall ~ k \in [0,K]},\forall ~ (i,j) \in E_{cc}, \label{eq:all:latency}\\
    &\sum_{(i,j)\in E_{cc}}\sum_{k=0}^{K}w_{i,j}*x_{uv}^{i,j,k}\leq BW_{max}, \forall (u,v)\in E_{\revise{link}}, \label{eq:s:bandwidth}\\
    &ip_{u}\leq max\_size, ~~~op_{u}\leq max\_size, \label{eq:s:port}\\
    &x^{i,j,k}_{uv}\in \{0,1\}, k \in [0,K], \forall(u,v)\in E_{\revise{pa}},\forall (i,j)\in E_{cc}. \label{eq:s:xBinary-all}
\end{align}
\end{subequations}

In the formulation above, the set of constraints (\ref{eq:s:unitFlow}) \minrev{defines} $K+1$ unit flows (paths) for each communication flow.
The constraint (\ref{eq:s:edgeDisjoint}) ensures that the $K+1$ paths of any communication flow $(i,j)$ are edge-disjoint.
The constraints (\ref{eq:all:latency}) ensure that the latency constraint is satisfied for \minrev{all $K+1$ routing paths} of each communication flow.
The set of constraints (\ref{eq:s:bandwidth}) \minrev{represents} the limited bandwidth of physical links and the set of constraints (\ref{eq:s:port}) \minrev{denotes} the maximum port number for each switch.

The objective (\ref{eq:ilpall}) is to minimize the total power consumption of switches and the total link power consumption of the default routing paths (k=0) for all communication flows.


The required runtime is unacceptable when directly solving the above large-scale ILP problem. To reduce the runtime of routing path allocation, we process the communication flows one by one in descending order of bandwidth requirements, which cause sub-optimal solutions but greatly reduce the runtime. Additionally, before processing each communication flow, the switch size constraint and the bandwidth constraint can be preprocessed by removing, from the graph $G_{pa}$, the \minrev{vertices} corresponding to the switches that have a maximum size and the edges corresponding to the links that have no enough \minrev{bandwidth}. Therefore, the ILP formulation for the routing path allocation of a single communication flow $(i,j)\in E_{cc}$ is simplified as follows.
\begin{subequations}
\label{eq:ilpsingle}
\allowdisplaybreaks[4]
\begin{align}
 Min~& \revise{\sum_{u\in V_s}P_{sw}(ip_u,op_u)} + \sum_{(u,v)\in E_{\revise{link}}}P_{\revise{lk}}(i,j,u,v) \label{eq:single:min} \tag{\ref*{eq:ilpsingle}}\\
    &\textrm{s.t.}\\
    &\sum_{v:(u,v)\in E_{\revise{pa}}}x_{uv}^{i,j,k} - \sum_{v:(v,u)\in E_{\revise{pa}}}x_{vu}^{i,j,k}= \nonumber
    \left\{
    \begin{array}{ll}
        1,      & \mbox{if}~ u=\revise{c_{i}}; \\
        0,      & \mbox{if}~ u\in \revise{V_{s}};\\
        -1.     & \mbox{if}~ u=\revise{c_{j}}; \\
    \end{array}\right.\nonumber\\
    &~~~~~~~~~~~~~~~\forall ~ k \in [0,K],  \\
    &\sum_{(u,u')\in E_{split}}x^{i,j,\minrev{k}}_{uu'}\leq l_{i,j}, \minrev{\forall ~ k \in [0,K]},\label{eq:s:latency}\\
    &x^{i,j,k}_{uv}\in \{0,1\}, k \in [0,K], \forall(u,v)\in E_{\revise{pa}}, \label{eq:s:xBinary-s}
\end{align}
\end{subequations}
where the cost of links $P_{\revise{lk}}(i,j,u,v)$ can be simply defined as follows.
\begin{equation}
\label{eq:cost-2}
P_{\revise{lk}}(i,j,u,v) = \begin{cases}
E_{bit}\times w_{i,j}\times D_{u,v}, ~if~physical~link~(u,v)~exists;\\
E_{bit}\times w_{i,j} \times D_{u,v} + C_{pl}, ~otherwise,\\
\end{cases}
\end{equation}


\subsubsection{An example}
\begin{figure}[htbp]
	\small \centering
	\includegraphics[width=9.00cm]{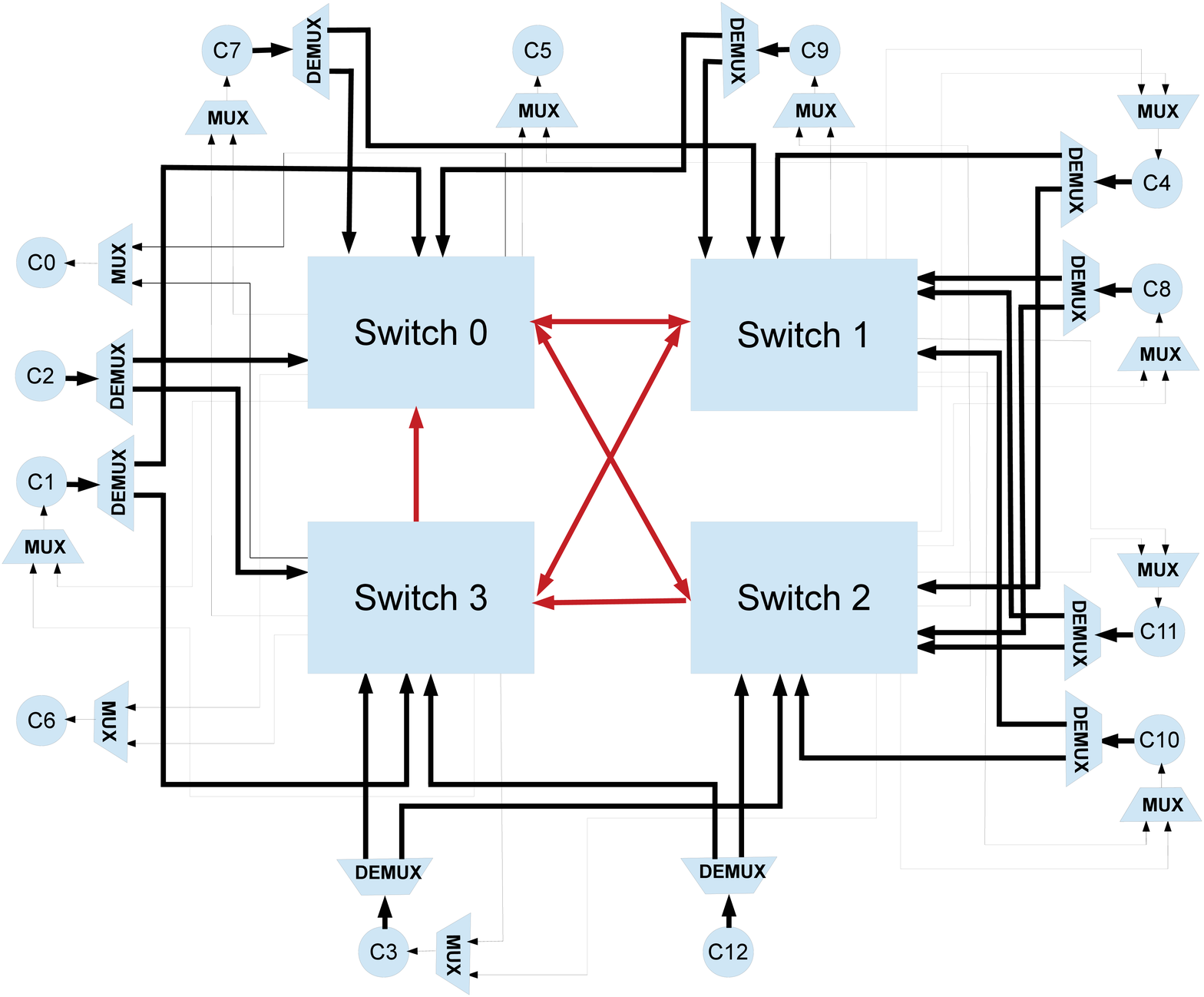}\\
	\caption{Final Network Topology.}
	\label{fig:resultOfPA}
\end{figure}

As mentioned above, we solve an ILP model for each communication flow $(c_i, c_j)$ in $G_{cc}$.
As shown in Fig.\ref{fig:resultOfPA}, we first allocate the routing paths of the communication flow $(c_3, c_0)$ and, a physical link from switch $s_{2}$ to $s_{0}$ is added, where the default path includes only one switch $s_{3}$ and the alternative path is from $s_{2}$ to $s_{0}$;
\minrev{After we allocate routing paths for} another communication flow $(c_4, c_3)$,
a physical link from switch $s_{1}$ to $s_{3}$ is added, where the default path includes only one switch $s_{2}$ and the alternative path is from $s_{1}$ to $s_{3}$.
Fig.\ref{fig:resultOfPA} shows the final network topology with one-fault tolerance, and the routing paths for all communication flows are shown in Table \ref{tab:sw:paths}.


\begin{table}[htbp]
	\centering
	\caption{Routing Paths for One-Fault Tolerance}
	\renewcommand{\arraystretch}{1.0}
	\newcommand{\tabincell}[2]{\begin{tabular}{@{}#1@{}}#2\end{tabular}}
	\begin{tabular}{ccc}
		\hline
		\textbf{Flows}& \textbf{Default Path} & \textbf{Alternative Path}\\
		\hline
		$c_{1}\rightarrow c_{0}$  &$s_{0}\rightarrow s_{0}$ &$s_{3}\rightarrow s_{3}$ \\
		$c_{2}\rightarrow c_{1}$  &$s_{0}\rightarrow s_{0}$ &$s_{3}\rightarrow s_{3}$ \\
		$c_{3}\rightarrow c_{0}$  &$s_{3}\rightarrow s_{3}$ &$s_{2}\rightarrow s_{0}$ \\
		$c_{4}\rightarrow c_{3}$  &$s_{2}\rightarrow s_{2}$ &$s_{1}\rightarrow s_{3}$ \\
		$c_{7}\rightarrow c_{5}$  &$s_{0}\rightarrow s_{0}$ &$s_{1}\rightarrow s_{1}$ \\
		$c_{7}\rightarrow c_{6}$  &$s_{0}\rightarrow s_{0}$ &$s_{1}\rightarrow s_{3}$ \\
		$c_{8}\rightarrow c_{7}$  &$s_{1}\rightarrow s_{0}$ &$s_{2}\rightarrow s_{3}$ \\
		$c_{9}\rightarrow c_{8}$  &$s_{1}\rightarrow s_{1}$ &$s_{0}\rightarrow s_{2}$ \\
		$c_{\minrev{10}}\rightarrow c_{\minrev{9}}$  &$s_{1}\rightarrow s_{\minrev{1}}$ &$s_{\minrev{2}}\rightarrow s_{\minrev{2}}$ \\
		$c_{11}\rightarrow c_{8}$  &$s_{1}\rightarrow s_{1}$ &$s_{2}\rightarrow s_{2}$ \\
		$c_{12}\rightarrow c_{4}$  &$s_{2}\rightarrow s_{2}$ &$s_{3}\rightarrow s_{1}$ \\
		$c_{12}\rightarrow c_{10}$  &$s_{2}\rightarrow s_{2}$ &$s_{3}\rightarrow s_{0}\rightarrow s_{1}$ \\
		$c_{12}\rightarrow c_{11}$  &$s_{2}\rightarrow s_{2}$ &$s_{3}\rightarrow s_{1}$ \\
		\hline
	\end{tabular}
	\label{tab:sw:paths}
\end{table}

\section{Switch Port Sharing}
\label{sec:ps}
In a $K$-fault tolerance structure, $K+1$ \textit{core inports} \minrev{or/}and $K+1$ \textit{core outports} are required for each core on no less than $K+1$ switches, which greatly increase the area and power consumption of switches.
Many switch ports are not used simultaneously because only one out of $K+1$ routing paths is used for each communication flow at a time.
To reduce the switch size, we propose \revise{the sharing of} switch ports (on the same switch) between the routing paths from different communication flows, using multiplexers. Given the routing path allocation of communication flows, we prove a sufficient and necessary condition for the port sharing on \minrev{a switch}, and formulated the problem into \minrev{a} clique partitioning problem.

The port sharing aims at the sharing of \textit{core inport}s/\textit{core outport}s on the same switch.
In this section, we first propose a two-stage method to solve the \textit{core inport} sharing problem, which is also applicable to \textit{core outport} sharing. \minrev{Second, we proposed a method to remove the conflicts caused by port sharing on multiple switches. Finally, an independent set based formulation is proposed for selecting routing paths for communication flows.}

\subsection{Conditions for Port Sharing \minrev{on a Switch}}
\label{sec:ps:SharingDecision}
\minrev{In this subsection, given a network topology with $K$-fault-tolerance and the corresponding routing path allocation, we derive the conditions for port sharing on one switch only.}
For clarity, the conditions for the inport sharing in one-fault-tolerance and $K$-fault-tolerance are respectively discussed in Section \ref{sec:ps:SharingDecision:Kone} and Section \ref{sec:ps:Inport:MultiplePaths}.

\subsubsection{Port Sharing in One-Fault-Tolerance Topologies}
\label{sec:ps:SharingDecision:Kone}

\begin{figure}[htbp]
\small \centering
  \includegraphics[width=8.50cm]{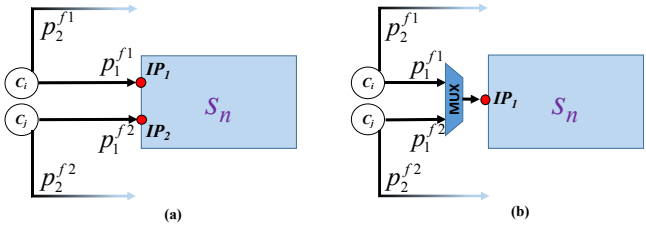}\\
  \caption{(a)Two independent core inports. (b) Port sharing.}
  \label{fig:sharingIntroduction}
\end{figure}

Suppose there are two core inports, $IP_{1}$ and $IP_{2}$, on a switch $s_{n}$, and two communication flows, $f_{1}$ and $f_{2}$ which are shown on the Fig.\ref{fig:sharingIntroduction}(a).
$f_{1}$  has two routing paths, $p^{f_1}_{1}$ and $p^{f_1}_{2}$, and $p^{f_1}_{1}$ goes through $IP_{1}$.  $f_{2}$ has two routing paths, $p^{f_2}_{1}$ and $p^{f_2}_{2}$, and $p^{f_2}_{1}$ goes through $IP_{2}$.
\revise{Fig.\ref{fig:sharingIntroduction}(b) \revise{illustrates} port sharing.}


\begin{lemma}
The two inports on $s_n$,  $IP_{1}$ and $IP_{2}$ respectively used by $p^{f_1}_{1}$ and $p^{f_2}_{1}$, can be merged into a single port without violating the property of one-fault tolerance if and only if $p^{f_1}_{2}$ and $p^{f_2}_{2}$ are vertex-disjoint.
\end{lemma}
Proof. \textit{IF.}  In the case that $p^{f_1}_{2}$ and $p^{f_2}_{2}$ are vertex-disjoint, after $IP_{1}$ and $IP_{2}$ are merged into a single port the routing paths for $f_{1}$ and $f_{2}$ can always be found when one fault occurs.
Because $p^{f_1}_{2}$ and $p^{f_2}_{2}$ cannot go through $s_n$, the one fault could be on $s_n$, $p^{f_1}_{2}$, or $p^{f_2}_{2}$.
If the fault is on switch $s_n$, $f_{1}$ and $f_{2}$ can use $p^{f_1}_{2}$ and $p^{f_2}_{2}$ for communication, respectively.
If the fault is on $p^{f_1}_{2}$ (or $p^{f_2}_{2}$), $f_{1}$ and $f_{2}$ can respectively use $p^{f_1}_{1}$ and $p^{f_2}_{2}$ (or $p^{f_1}_{2}$ and $p^{f_2}_{1}$) for communication. It is concluded that $p^{f_1}_{1}$ and $p^{f_2}_{1}$, respectively going through $IP_{1}$ and $IP_{2}$, are not used simultaneously while the topology keeps the property of one-fault tolerance.

\textit{ONLY IF.} We show that if $p^{f_1}_{2}$ and $p^{f_2}_{2}$ are not vertex-disjoint, merging $IP_{1}$ and $IP_{2}$ will violate the one-fault tolerance.  $p^{f_1}_{2}$ and $p^{f_2}_{2}$ must go through one common switch vertex if they are not vertex-disjoint. Consequently, if the common switch is faulty then $f_{1}$ and $f_{2}$ have to use $p^{f_1}_{1}$ and $p^{f_2}_{1}$ for communication, which causes conflict use of core inports.
\revise{\hfill \textbf{Proof END}.}

\begin{figure}[htbp]
\small \centering
  \includegraphics[width=7.0cm,height=3.0cm]{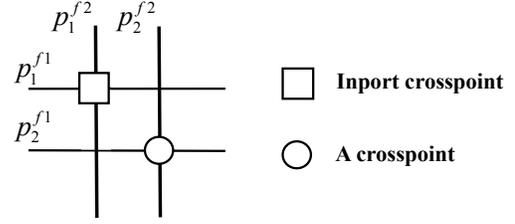}\\
  \caption{The intersection relation violates the one-fault tolerance considering the routing paths of $f_{1}$ and $f_{2}$.}
  \label{fig:possibleCrossPathOfInport:Kone}
\end{figure}

Fig.\ref{fig:possibleCrossPathOfInport:Kone} shows the intersection relations between the routing paths of the communication flows $f_1$ and $f_2$, which causes core inport conflict if $IP_1$ and $IP_2$ are merged and one fault occurs in the intersection point of routing paths $p^{f_1}_{2}$ and $p^{f_2}_{2}$.


\subsubsection{Port Sharing in Multiple-Fault Tolerance Topologies}

In this subsection, we give a generalized sufficient and necessary condition for port sharing in $K$-fault-tolerant ($K\ge 1$) structures.

Suppose there are two core inports, $IP_{1}$ and $IP_{2}$, on a switch $s_{n}$, and two communication flows, $f_{1}$ and $f_{2}$. $f_{1}$ has $K+1$ vertex-disjoint routing paths, $p^{f_1}_{0},~p^{f_1}_{1},\cdots,~p^{f_1}_{K}$, of which $p^{f_1}_{0}$ goes through $IP_1$.  $f_{2}$ has $K+1$ vertex-disjoint routing paths, $p^{f_2}_{0},~p^{f_2}_{1},\cdots,~p^{f_2}_{K}$, of which $p^{f_2}_{0}$ goes through $IP_2$.

We construct a bipartite graph, $IG(V, E)$, to represent the intersection relations between the routing paths of $f_{1}$ and $f_{2}$ as follows.

The vertex set includes all the routing paths of $f_{1}$ and $f_{2}$ except for $p^{f_1}_{0}$ and $p^{f_2}_{0}$. Let $P_{f_1} = \{p^{f_1}_{i}, i=1, \cdots, K\}$ and $P_{f_2}=\{p^{f_2}_{i}, i=1, \cdots, K\}$.  $V_{IG} =  P_{f_1} \cup  P_{f_2} $. \revise{Further,} if two routing paths from $P_{f_1}$  and $P_{f_2}$ have a common vertex, there is an edge in $E_{IG}$. That is,  $E_{IG} = \{ (p^{f_1}_{i}, p^{f_2}_{j})| p^{f_1}_{i}\in P_{f_1}~p^{f_2}_{j} \in P_{f_2}~and~they~have~ a~ common~ vertex.\}$.  It is obvious that $IG(V, E)$ is a bipartite graph. Let $C(IG)$ be the maximum cardinality matching in $IG(V,E)$. \revise{Then,} we have the following conclusion.

\begin{theorem}
\label{theorem:ps:OneFlow}
The two core inports on $s_n$,  $IP_{1}$ and $IP_{2}$, respectively used by $p^{f_1}_{0}$ of $f_1$ and $p^{f_2}_{0}$ of $f_2$, can be merged into a single port (port sharing) without violating the property of $K$-fault tolerance if and only if $C(IG)< K$.
\end{theorem}
\textbf{Proof.} \textit{IF.} We show that if $C(IG)< K$ the merging of $IP_{1}$ and $IP_{2}$ will not violate the $K$-fault tolerance.
Note that the routing paths from $P_{f_1}$ (or $P_{f_2}$) are vertex-disjoint  (except for the source and sink core \minrev{vertices}).
One fault causes at most two faulty paths, of which one path is from $P_{f_1}$ and the other is from $P_{f_2}$ and they have a common vertex.
We have two situations considering $p^{f_1}_{0}$ and $p^{f_2}_{0}$.

(1). $p^{f_1}_{0}$ and $p^{f_2}_{0}$ are correct but only one of them can be used since they share one core inport. Let $K_1$ and $K_2$ respectively be the number of faulty paths in $P_{f_1}$ and $P_{f_2}$. Without loss of generality, we suppose $K_1\ge K_2$. When $K_1 < K$, there must be at least a correct routing path in both $P_{f_1}$ and $P_{f_2}$ and, hence, $K$ faults are tolerant. When $K_1 = K$, we can conclude that $K_2 < K$. Because $K_2 = K_1=K$ indicates that $C(IG) = K$, which contradicts $C(IG)< K$. Accordingly, we have at least a correct path in $P_{f_2}$ for communication flow $f_2$ and $p^{f_1}_{0}$ can be used for $f_1$. Consequently, the network topology \revise{maintains} $K$-fault tolerance.


(2).  $p^{f_1}_{0}$ and $p^{f_2}_{0}$ are faulty when $s_n$ is faulty. In this case, the paths in $P_{f_1}$ and $P_{f_2}$ are able to construct a $K-1$ fault-tolerance structure since the given topology is $K$-fault tolerant. Consequently, the network topology also keeps $K$-fault tolerance.

\textit{ONLY IF.}
We show that if $C(IG) = K$, then merging $IP_{1}$ and $IP_{2}$ will violate the $K$-fault tolerance.
There will be perfect matching in $IG(V,E)$ if $C(IG) = K$.
In the perfect matching, each edge corresponds to an intersection of two routing paths from $P_{f_1}$ and $P_{f_2}$.
All $2K$ paths in $P_{f_1}$ and $P_{f_2}$ are faulty when $K$ faults exactly occur on $K$ intersected \minrev{vertices}. Consequently,  $p^{f_1}_{0}$ and $p^{f_2}_{0}$ are respectively the only available routing path for $f_1$ and $f_2$, which cause a conflicting use of merged core inports if we merge $IP_1$ and $IP_2$. \revise{\hfill \textbf{Proof END}.}

 \begin{figure}[htbp]
\small \centering
  \includegraphics[width=8.50cm]{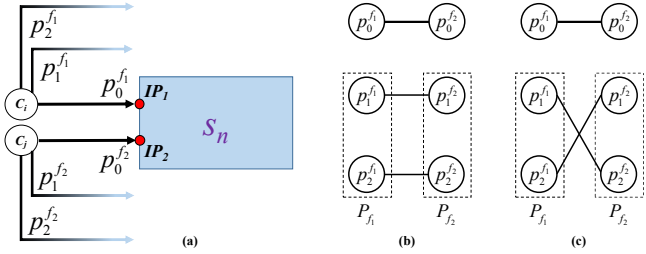}\\
  \caption{The intersection relation violates the 2-fault tolerance considering the routing paths of $f_{1}$ and $f_{2}$ where $p^{f_{1}}_{1}$ intersects $p^{f_{2}}_{1}$ at router $s_{n}$.}
  \label{fig:BinaryGraphOfMultiplePaths}
\end{figure}
Further, Fig.\ref{fig:BinaryGraphOfMultiplePaths} shows the intersection relation violates the 2-fault tolerance considering the routing paths of $f_{1}$ and $f_{2}$ where $p^{f_{1}}_{1}$ intersects $p^{f_{2}}_{1}$ at router $s_{n}$.

\subsubsection{Inport with Multiple Communication Flows}
\label{sec:ps:Inport:MultiplePaths}

For each core inport, there may be one or one more outgoing communication requirements. Accordingly, there may be one or multiple paths \revise{going} through one core inport.

\begin{figure}[htbp]
\small \centering
  \includegraphics[width=6.00cm]{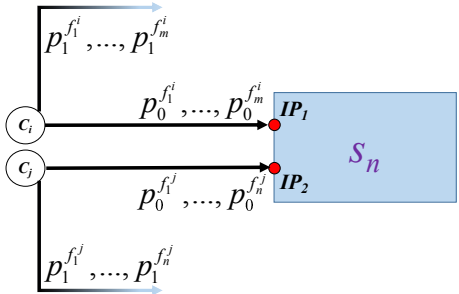}\\
  \caption{Multiple paths of the inports.}
  \label{fig:inportBeforeSharingMultiPaths}
\end{figure}

Suppose there are two core inports, $IP_{1}$ and $IP_{2}$, on a switch $s_{n}$ and there are respectively $m$ and $n$ communication flows going through $IP_{1}$ and $IP_{2}$,  as shown in Fig.\ref{fig:inportBeforeSharingMultiPaths}.

\begin{corollary}
\label{corollary:ps:MultiFlow}
The two inports $IP_{1}$ and $IP_{2}$ can be merged if and only if all the $m\times n$ pairs of communication flows satisfy Theorem \ref{theorem:ps:OneFlow}.
\end{corollary}


\subsubsection{Multiple Port Sharing}
\label{sec:ps:MultipleInportSharing}

Suppose there are $J$ core inports, $IP_{1}, IP_{2}, \cdots, IP_{J}$, on a switch $s_{n}$, and $J$ communication flows, $f_{1}, f_{2}, \cdots, f_{J}$. $f_{i}, i=1, \cdots, J$, has $K+1$ vertex-disjoint routing paths, $p^{f_i}_{0},~p^{f_i}_{1},\cdots,~p^{f_i}_{K}$, of which $p^{f_i}_{0}$ goes through $IP_i$.

\begin{theorem}
\label{theorem:ps:MultiPortsSharing}
The $J$ core inports on $s_n$,  $IP_j, j=1, \cdots, J$, respectively used by $J$ communication flows, $f_{1}, f_{2}, \cdots, f_{J}$, can be merged into a single inport without violating the property of $K$-fault tolerance if the merging relations between $J$ inports form a clique, which indicates that, for all pairs of $(IP_i, IP_j)$, $i\ne j$, and $1\le i,j\le J$, $IP_i$ and $IP_j$ can be merged according to Theorem \ref{theorem:ps:OneFlow}.
\end{theorem}
\textbf{Proof.}
Let $P_{f_j} = \{p^{f_j}_{k}, k=1, \cdots, K\}$ be the set of vertex-disjoint routing paths of $f_j$ except for $p^{f_j}_{0}$.
Similarly, we have two situations considering $p^{f_i}_{0},~i=1, \cdots, J$.

(1). $p^{f_j}_{0},~j=1, \cdots, J$ are correct but only one of them can be used since they share one core inport.
Let $K_j,~j=1, \cdots, J$, respectively be the number of faulty paths in $P_{f_j}$.
Without loss of generality, we suppose that $K_j\ge K_{j+1}, j=1, \cdots, J-1$.
When $K_1 < K$, there must be at least a correct routing path in $P_{f_j}, j=1, \cdots, J$, and, hence, $K$-faults are tolerant.
When $K_1 = K$, we can conclude that $K_j < K, j=2, \cdots, J$.
\revise{This is because} $K_j = K_1=K$ indicates that there will be perfect matching considering the intersection relations between the $K$ paths from $P_{f_1}$ and $K$ paths from $P_{f_j}$ and, hence, $IP_1$ and $IP_j$ cannot be merged according to Theorem \ref{theorem:ps:OneFlow}, which is a contradiction.
Accordingly, we have at least a correct path in $P_{f_j}$ for communication flow ${f_j}$, $j=2, \cdots,~J$ and $p^{f_1}_{0}$ can be used for $f_1$.
Consequently, the network topology \revise{maintains} $K$-fault tolerance.

(2). $p^{f_j}_{0},~j=1, \cdots, J$ are faulty when $s_n$ is faulty.
In this case, the paths in $P_{f_j}, j=1, \cdots, K$, are able to construct a $K-1$ fault-tolerance structure since the given topology is $K$-fault tolerant.
Consequently, the network topology also \revise{maintains} $K$-fault tolerance.\revise{\hfill \textbf{Proof END}.}

Suppose there are $J$ core inports, $IP_{1}, IP_{2}, \cdots, IP_{J}$, on a switch $s_{n}$, and there are $m_i$  communication flows going through $IP_{i}$, $i=1,\cdots, J$.

\begin{corollary}
\label{corollary:ps:MultiFlowMultiPort}
The $J$ core inports on $s_n$,  $IP_j, j=1, \cdots, J$, respectively used by  \revise{$m_j,  j=1, \cdots, J,$ communication flows,} can be merged into a single inport without violating the property if all the $\prod_{i=1}^{J}{m_i}$ combinations of communication flows satisfy Theorem \ref{theorem:ps:MultiPortsSharing}.
\end{corollary}

Based on the above theorem and corollaries, we conclude the following theorem.

\begin{theorem}
\label{theorem:ps:MP-ML-Sharing}
The $J$ core inports on $s_n$,  $IP_j, j=1, \cdots, J$, respectively used by $m_j,  j=1, \cdots, J$,  communication flows, can be merged into a single inport without violating the property of $K$-fault tolerance if the merging relations between $J$ inports form a clique.
\end{theorem}

\subsection{Clique Partitioning for Port Sharing on \minrev{a switch}}
\label{sec:ps:MaximumSharing}

Given a network topology \minrev{and all routing paths for $K$-fault-tolerance, we formulate the port sharing problem on a switch as a clique partitioning problem, where the clique number is minimized to reduce the switch size, according to Theorem \ref{theorem:ps:MP-ML-Sharing}}.

\begin{figure}[htbp]
\small \centering
  \includegraphics[width=8cm]{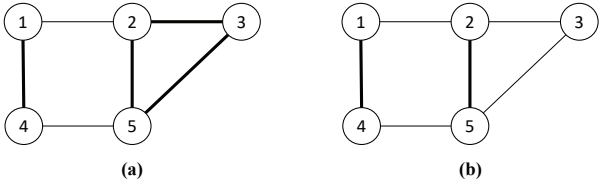}\\
  \caption{An example of $G_{ps}$.}
  \label{fig:PossibleSharingGraphTest}
\end{figure}

For a switch, a graph $G_{ps}(V_{ps}, E_{ps})$ is constructed to represent the possible sharing relations between the core inports.
The vertex set $V_{ps} = \{ IP_i, i=0, \cdots, N\}$ represents the set of core inports. An edge, $(IP_i, IP_j)$, is added to $E_{ps}$ if $IP_i$ and $IP_j$ can be shared with each other according to Corollary \ref{corollary:ps:MultiFlow}.
Fig.\ref{fig:PossibleSharingGraphTest} shows an example of $G_{ps}$ and its two clique-partitioning. The solid edges are the clique edges. In Fig.\ref{fig:PossibleSharingGraphTest}.(a), two inports are required, respectively corresponding to a 2-vertex clique and 3-vertex clique. In Fig.\ref{fig:PossibleSharingGraphTest}.(\minrev{b}), three inports are required, respectively corresponding to two 2-vertex cliques and one 1-vertex clique.

\minrev{Because} the clique partitioning problem is an NP-hard problem, we propose a heuristic to find the clique partitioning of $G_{ps}$.
\textbf{Algorithm} $1$ shows the key steps of \minrev{the heuristic for} port sharing.
Firstly, we find a maximum clique $Q_{V_{q},E_{q}}$ in $G_{ps}$ using an ILP based method \cite{pardalos1992branch}.
If $|V_{q}|$ is greater than $2$, we merge the inports in $V_{q}$ into a single port and remove $Q_{V_{q},E_{q}}$ from $G_{ps}$ to update a new $G_{ps}$.
The operation of finding the maximum clique is repeated until $|V_{q}| \le 2$,  where the clique partitioning problem can be solved by finding maximum cardinality matching.
Secondly, we find maximum cardinality matching \revise{in the rest of} $G_{ps}$, where the edges in the maximum cardinality matching correspond to two-port sharing.

\begin{algorithm}
\label{alg:portSharing}
\caption{\textit{\minrev{PS\_on\_a\_switch ($s_n$)}}}
\begin{algorithmic}
\Require{fault tolerant paths of communication requirements}
\Ensure{results of port sharing \minrev{within a switch}}
  \State Construct a possible port-sharing graph $G_{ps}$ for $s_{n}$;
    \For{each pair of inports(or outports) ($IP_{i}$,$IP_{j}$) on $s_{n}$}
    \For{each path $p^{f^{i}_{m}}_{1}$ through $IP_{i}$}
            \For{each path $p^{f^{j}_{n}}_{1}$ through $IP_{j}$}
            \State Construct a bipartite graph $IG$;
            \State Calculate $C(IG)$ of $IG$;
                \If{$C(IG)$ $>=$ $K$}
                    \State{\textbf{goto} \revise{cont};}
                \EndIf
            \EndFor
        \EndFor
        \State Add an edge $(IP_{i},IP_{j})$ to $E_{gs}$;\\
    \revise{cont: \minrev{Continue}};
    \EndFor
    \Repeat
    \State Find \minrev{a maximum clique} $Q(V_{q},E_{q})$ in $G_{ps}$ and the IPs denoted by $V_{q}$ share a single port;
    \State Remove $Q(V_{q},E_{q})$ from $G_{ps}$;
    \Until{$|V_{q}| \le 2$}
    \State \minrev{Find maximum cardinality matching on $G_{ps}$;}
\end{algorithmic}
\end{algorithm}

\begin{color}{blue}

\subsection{Port Sharing on Multiple Switches}
\label{subsec:psms}
The conditions in Section \ref{sec:ps:SharingDecision} ensure the fault tolerance when port sharing is considered on one switch only.
However, port sharing on multiple switches perhaps causes conflicts of routing path selection. In this section, we present a method for removing some port sharing to maintain the fault tolerance when port sharing on all switches are considered.
\end{color}

To select routing paths for communication flows, we can construct a graph $G_{pc}(V_{pc},E_{pc})$ to represent the conflict relations between routing paths and solving an independent set problem on $G_{pc}$.
Let $p^k_{i,j}$ be the $k$-th routing path of communication flow $(i,j)$.
$V_{pc} =\{p^k_{i,j}| (i,j)\in E_{cc}, 0\le k \le K\}$ represent the routing paths of all communication flows.
$E_{pc}$ includes two types of edges. One represents two routing paths from the same communication flow and the other represents port-sharing relations between two routing paths.
$E_{pc}= \{ (p^{k_1}_{(i,j)}, p^{k_2}_{(i,j)})| 0\le k_1,k_2 \le K \text{ and } k_1\neq k_2 \} \cup \{ (p^{k_1}_{(i_1,j_1)}, p^{k_2}_{(i_2,j_2)})|~(i_1,j_1)\neq (i_2,j_2), 0\le k_1,k_2 \le K$, and $p^{k_1}_{(i_1,j_1)}$ and $p^{k_2}_{(i_2,j_2)}$  go through a common \textit{core inport} or \textit{core outport}\}.

\begin{color}{blue}
The selection of routing paths can be achieved by finding a maximum independent set of $G_{pc}$, denoted as $IND(G_{pc})$. If there is no port sharing, that is, there is no second type of edges in $E_{pc}$, we may choose any correct routing path for each communication flow, and, hence  $|IND(G_{pc})| = |E_{cc}|$. If port sharing is considered on one switch only, the conditions in Section \ref{sec:ps:SharingDecision} ensure $|IND(G_{pc})| = |E_{cc}|$ for any $K$ switch faults. However, $|IND(G_{pc})| < |E_{cc}|$ could happen if we have two or more switches with shared \textit{core inports/outports}, that is, we cannot find enough routing paths for the communication flows.

Fig.\ref{fig:conflict-exmaple} shows an example. We have three communication flows $f_1$, $f_2$, and $f_3$ ($|E_{cc}|=3$), and each flow has two switch-disjoint routing paths, $p^{f_i}_1$ and $p^{f_i}_2$, for one-fault tolerance. As shown in the figure, the routing paths have two shared ports respectively on the switches $S_m$ and $S_n$, and $p^{f_1}_2$ and $p^{f_3}_2$ go through a common switch $S_l$. The corresponding $G_{pc}$ is also shown in the figure. When $s_l$ is broken, we have four correct paths after removing $p^{f_1}_2$ and $p^{f_3}_2$. Obviously, $|IND(G_{pc})| = 2 < 3 = |E_{cc}|$, which cause that one flow has no routing paths. 
The conflict can be solved by removing the port sharing on any switch.
\begin{figure}[htbp]
\small \centering
  \includegraphics[width=8.50cm]{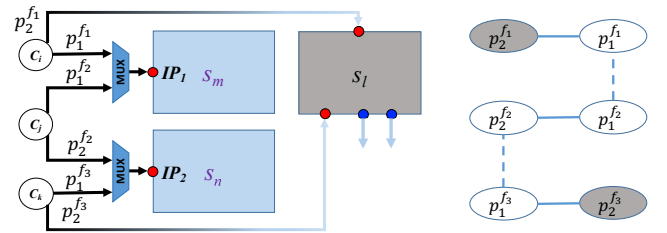}\\
  \caption{Example for the conflicts of routing paths with port sharing on two switches.}
  \label{fig:conflict-exmaple}
\end{figure}

Here, we propose a heuristic to deal with conflicts of port sharing for all the switches. Let $V_f$ be the set of faulty routers. Algorithm $2$ shows the key steps.

\begin{algorithm}
\label{alg:portSharingall}
\caption{\textit{\minrev{PS\_on\_multiple\_switches}}}
\begin{algorithmic}
\Require{fault tolerant paths of communication requirements}
\Ensure{results of port sharing}
\For{each switch $s_{n}$ in $V_s$}
  \State Call  \textit{PS\_on\_a\_switch($s_n$)} to generate \textit{core inport} sharing;
  \State Call  \textit{PS\_on\_a\_switch($s_n$)} to generate \textit{core outport} sharing;
\EndFor
\State Construct a conflict relation graph $G_{pc}$ between routing paths;
\For{each subset of switches $V_f\subset V_s$ and $|V_f| = K$}
  \State Temporarily remove from $G_{pc}$ the routing paths going through any switch $s_i \in V_f$;
  \State Find a maximum independent set of $G_{pc}$;
  \If{$|IND(G_{pc})| = |E_{cc}|$}
     continue;
  \EndIf
  \Repeat
    \State Choose a communication flow $(i, j)$ having no routing path in $IND(G_{pc})$;
    \State Permanently remove one of the port-sharing edges related to $(i,j)$. 
    \State Find a maximum independent set of $G_{pc}$;
  \Until{$|IND(G_{pc})| = |E_{cc}|$}
\State Restore $G_{pc}$, except for the permanently removed edge;
\EndFor
\end{algorithmic}
\end{algorithm}

In Algorithm $2$, we first generate the \textit{core inport/outport} sharing for each switch by calling Algorithm $1$.
Second, for each subset of possible $K$ faulty switches, the fault-tolerance is verified by solving a maximum independent set problem, where $|IND(G_{pc})|$ is computed by solving an ILP-based formulation. When $|IND(G_{pc})| < |E_{cc}|$, a port-sharing edge is considered to be removed for increasing the size of maximum independent set.
Basically, we consider removal of a port-sharing edge between \textit{core outport}s while keeping the sharing edges between \textit{core inport}s as many as possible, because an input port generally has a flit buffer with large area costs and power overhead.

\end{color}

\begin{figure}[htbp]
\small \centering
  \includegraphics[width=8.50cm]{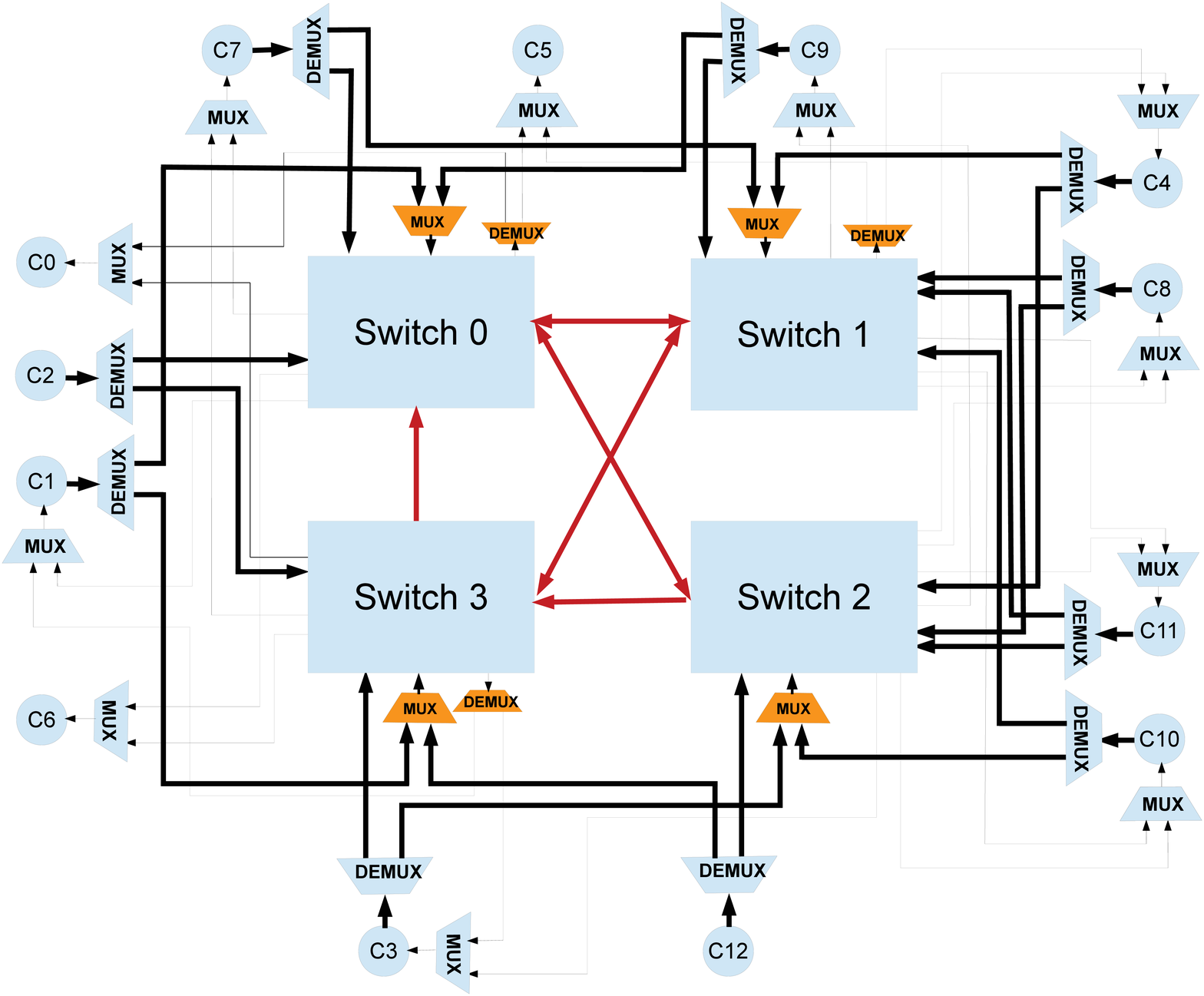}\\
  \caption{Result of Port Sharing.}
  \label{fig:resultOfSharing}
\end{figure}

\begin{figure}[htbp]
\small \centering
  \includegraphics[width=7.0cm]{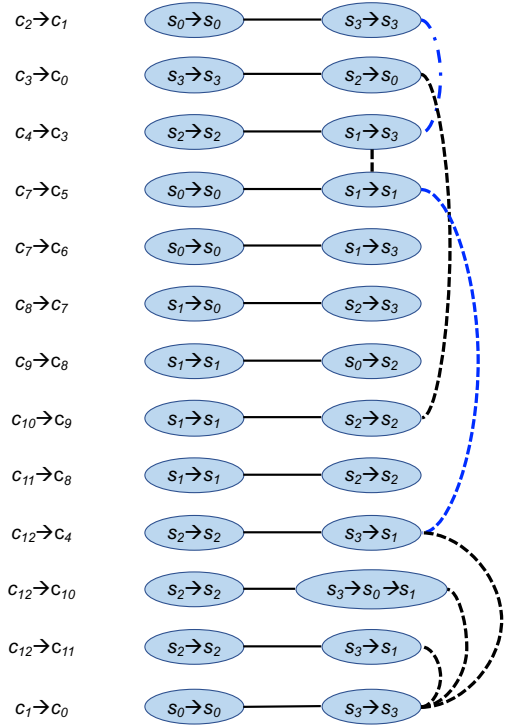}\\
  \caption{\minrev{$G_{pc}$ for selecting routing paths in Fig.\ref{fig:resultOfSharing}.}}
  \label{fig:gpc-example}
\end{figure}


\begin{table}[htbp]
	\centering
	\caption{Communication paths for one-switch-fault tolerance with $s_{0}$ broken down}
	\definecolor{gray}{RGB}{140,140,140}
	\renewcommand{\arraystretch}{1.0}
	\addtolength{\tabcolsep}{-1.1pt}
	\newcommand{\tabincell}[2]{\begin{tabular}{@{}#1@{}}#2\end{tabular}}
	\begin{tabular}{ccc}
		\hline
		\textbf{Flows}& \textbf{Default Path} & \textbf{Alternative Path}\\
		\hline
		$c_{1}\rightarrow c_{0}$  &\textcolor{gray}{$s_{0}\rightarrow s_{0}$} &$s_{3}\rightarrow s_{3}$ \\
		$c_{2}\rightarrow c_{1}$  &\textcolor{gray}{$s_{0}\rightarrow s_{0}$} &$s_{3}\rightarrow s_{3}$ \\
		$c_{3}\rightarrow c_{0}$  &$s_{\minrev{3}}\rightarrow s_{\minrev{3}}$ &\textcolor{gray}{$s_{\minrev{2}}\rightarrow s_{\minrev{0}}$ }\\
		$c_{4}\rightarrow c_{3}$  &$s_{2}\rightarrow s_{2}$ &\textcolor{gray}{$s_{1}\rightarrow s_{3}$}\minrev{[c]}\\
		$c_{7}\rightarrow c_{5}$  &\textcolor{gray}{$s_{0}\rightarrow s_{0}$} &$s_{1}\rightarrow s_{1}$ \\
		$c_{7}\rightarrow c_{6}$  &\textcolor{gray}{$s_{0}\rightarrow s_{0}$} &$s_{1}\rightarrow s_{3}$ \\
		$c_{8}\rightarrow c_{7}$  &\textcolor{gray}{$s_{1}\rightarrow s_{0}$} &$s_{2}\rightarrow s_{3}$ \\
		$c_{9}\rightarrow c_{8}$  &$s_{1}\rightarrow s_{1}$ &\textcolor{gray}{$s_{0}\rightarrow s_{2}$} \\
    $\minrev{c_{10}}\rightarrow \minrev{c_{9}}$  &$\minrev{{s_{1}}}\rightarrow \minrev{s_{1}}$ &$\minrev{s_{2}}\rightarrow \minrev{s_{2}}$ \\
		$c_{11}\rightarrow c_{8}$  &$s_{1}\rightarrow s_{1}$ &$s_{2}\rightarrow s_{2}$ \\
		$c_{12}\rightarrow c_{4}$  &$s_{2}\rightarrow s_{2}$ &\textcolor{gray}{$s_{3}\rightarrow s_{1}$}\minrev{[c]} \\
		$c_{12}\rightarrow c_{10}$  &$s_{2}\rightarrow s_{2}$ &\textcolor{gray}{$s_{3}\rightarrow s_{0}\rightarrow s_{1}$} \\
		$c_{12}\rightarrow c_{11}$  &$s_{2}\rightarrow s_{2}$ &\textcolor{gray}{$s_{3}\rightarrow s_{1}$}\minrev{[c]} \\
		\hline
	\end{tabular}
	\label{tab:sw:pathsAfterFailure}
\end{table}

Fig.\ref{fig:resultOfSharing} shows the port sharing of the network topology in Fig. \ref{fig:resultOfPA}; the corresponding $G_{cc}$ in Fig.\ref{fig:mp3CCG}.
\minrev{Fig.\ref{fig:gpc-example} shows the graph $G_{pc}$ corresponding to Fig.\ref{fig:resultOfSharing}.}
When the switch $s_{0}$ is broken down, Table \ref{tab:sw:pathsAfterFailure} shows \minrev{the available routing paths}, where the faulty paths are displayed in gray color and \minrev{the conflict paths are displayed in gray color and marked using $c$.}

\subsection{Selecting Routing Paths after Port Sharing}
\label{subsec:pathsel}



The selection of routing paths can be achieved by finding an $|E_{cc}|$-size independent set of $G_{pc}$.
If link faults or switch faults occur, we can just remove the routing paths that go through the faulty links and the faulty switches from $G_{pc}$ and find a new set of routing paths for all the communication flows by solving an $|E_{cc}|$-size independent set problem on $G_{pc}$ and update the routing tables of the core communications.

\subsection{Cost Analysis for Multiplexers and Demutiplexers}
\label{subsec:costa}
\revise{To develop a fault-tolerance topology}, we introduce demultiplexers ($DEMUX$) and multiplexers ($MUX$) for the source cores and the sink cores, respectively, of the communications flows. \revise{The routing paths can be selected by sending to the $DEMUX$s and $MUX$s control signals.}

\revise{Because we assume source-routing strategy, where the routing paths are stored in a routing table in the source core side of the communication flow. Each digit of the routing information is used in turn to select the output port at each step of the route, as if the address itself was the routing header determined from a source-routing table \cite{noc-stanf}.
Hence, for a demultiplexer with the input from a core (for example, the $DEMUX$ connected to $c7$ in Fig.\ref{fig:resultOfSharing}), the control signals can be from the routing bits, and for a demultiplexer with the input from a switch (for example, the $DEMUX$ connected to $switch~1$ in Fig.\ref{fig:resultOfSharing}), the control signals can also be generated by the switch according to the routing digit in the head flit of packet.}
\revise{For a multiplexer, we can send one-bit enable signal along with each input data, and the control signals of the multiplexer can be generated based on the enable signals using a simple logic circuit, which includes several logic gates. For $K$-fault tolerance, we need $K+1$ enable signals. In practical designs, $K$ will be very small ($K\le 3$ in this work), and hence, the power and area overhead is very small. Notice that the $MUX$ and $DEMUX$ only exist at the starting point and ending point of routing paths.}

\revise{In the following, we analyze the costs of $DEMUX$ and $MUX$.}
We set the bit-width to 32 bits and synthesize $DEMUX$ and $MUX$ with different sizes based on the $65nm$ process technology using commercial logic synthesis tools. The power consumption is shown in Table \ref{tab:basic:MUXandDEMUX}.
From the table, we can see that the power consumption of $DEMUX$ and $MUX$ is at least two order of magnitude less than that of the switch.
Consequently, the power consumption from introducing one more $DEMUX$ and $MUX$ port is much less than the power consumption reduced by \revise{removing} a switch port.
\begin{table}[htbp]
	\centering
	\caption{Power consumption of $DEMUX$ and $MUX$}
	\begin{tabular}{cccc}
		\hline
		\textbf{Size} & \textbf{Leakage power($\mu W$)} &\textbf{Total Power($mW$)} \\
		\hline
		$MUX\_2:1$ &1.1269e-02	&2.2274e-03\\
		$MUX\_3:1$ &1.2971e-02	&2.9082e-02\\
		$MUX\_4:1$ &2.1191e-02	&5.3280e-02\\
		$MUX\_5:1$ &2.6081e-02	&6.3278e-02\\
		$MUX\_6:1$ &3.0442e-02	&7.1425e-02\\
		\hline
		\hline
		$DEMUX\_1:2$ &1.9218e-02	&1.8354e-03\\
		$DEMUX\_1:3$ &2.6286e-02	&3.4385e-02\\
		$DEMUX\_1:4$ &1.6736e-02	&4.8588e-02\\
		$DEMUX\_1:5$ &1.5607e-02	&5.0520e-02\\
		$DEMUX\_1:6$ &1.7771e-02	&6.0453e-02\\
		\hline
	\end{tabular}
	\label{tab:basic:MUXandDEMUX}
\end{table}


\section{Link-Fault Tolerance} \label{sec:linkonly}

In this section, we consider the generation of $K$-link-fault-tolerance network topology, which is a special case of the generalized fault-tolerance topology.
If only link failures between switches are considered, the mapping from cores to switches \revise{exhibits} a many-to-one relationship. 
We propose an ILP based method to simultaneously solve the core mapping problem and the routing path allocation problem to improve the quality of the solutions.

Here, we define a routing path graph $G_{rp}(V_{rp}, E_{rp})$ to represent the possible connections between the cores and the switches and the possible physical links between the switches.
\noindent\definition{Routing Path Graph: $G_{rp}(V_{rp}, E_{rp})$ is directed, and $V_{rp}=V_{c} \cup V_{s}$ and $E_{rp}= \minrev{V_s\times V_s \cup V_{c}\times V_{s} \cup V_{s} \times V_{c}}$.

\revise{In the following, we give an ILP formulation  to find $K + 1$ edge-disjoint routing paths in $G_{rp}$ for all the communication flows. The objective is to minimize the power consumption of the NoC topology with a switch size constraint, bandwidth constraints, and latency constraints.
}

The initial number of switches $n_{sw}$ is determined using method \revise{similar to the one} in \cite{FT}. \revise{The binary variables $x^{(i,j,k)}_{uv}$} and \textbf{$d_{uv}$} are defined similar to those in the formulation (\ref{eq:ilpall}).
The $K$-link-fault-tolerant topology generation problem can be formulated as the following integer programming problem:
\begin{subequations}
	\label{eq:ilpLink}
	\allowdisplaybreaks[4]
	\begin{align}
	Min~~ & \sum_{u\in V_s}P_{\revise{sw}}(ip_u,op_u) + \revise{P_{link}} \tag{\ref*{eq:ilpLink}}\\
	\textrm{s.t.} \ \
	&\sum_{v:(u,v)\in E_{rp}}x^{(i,j,k)}_{uv} - \sum_{v:(u,v)\in E_{rp}}x^{(i,j,k)}_{vu} = \nonumber \\
	&\quad \left\{
	\begin{array}{ll}
	1,         & \mbox{if}~ u=c_{i}; \\
	0,         & \mbox{if}~ u\in V_{s}; \\
	-1,        & \mbox{if}~ u=c_{j}; \\
	\end{array} \right.           \forall (i,j)\in E_{cc},k\in [0,K]   \label{eq:link:unitFlow}\\
	&\sum_{k\in [0,K]}x^{(i,j,k)}_{uv}\leq 1, \forall (u,v)\in E_{rp}, (i,j)\in E_{cc} \label{eq:link:disjoint-edge}\\
	&\sum_{u\in V_{c}}d_{uv}\leq max\_size-1, \forall v\in V_{s} \label{eq:link:corePort}\\
	&ip_{u}\leq max\_size, ~~~op_{u}\leq max\_size\label{eq:link:allPort}\\
	&\sum_{(u,v)\in \minrev{V_s\times V_s}}x^{(i,j,0)}_{uv}+1\leq l_{i,j},  \forall (i,j)\in E_{cc}, \label{eq:link:latency}\\
	&\sum_{(i,j)\in E_{cc},k\in [0,K]}x_{uv}^{(i,j,k)}\cdot w_{i,j}\leq BW_{max}, \forall (u,v)\in V_{rp} \label{eq:link:bandwidth}\\
	&\sum_{v:(u,v)\in \minrev{V_c\times V_s}}d_{uv} = 1, \forall u\in V_{c} \label{eq:link:coreLinkRouter}\\
		&\sum_{v:(v,u)\in \minrev{V_s\times V_c}}d_{vu} = 1, \forall u\in V_{c} \label{eq:link:routerLinkcore}\\
	&x^{(i,j,k)}_{uv}\in \{0,1\}, \forall (u,v)\in E_{rp},(i,j)\in E_{cc},k\in [0,K] \label{eq:link:xBinary}\\
	&d_{uv}\in \{0,1\}, \forall (u,v)\in E_{rp} \label{eq:link:duvBinary}
	\end{align}
\end{subequations}

\revise{$P_{\revise{sw}}(ip_u,op_u)$ and $P_{link}$ are computed using a method similar to the one in Section \ref{subsec:power-model}.}
The constraint (\ref{eq:link:unitFlow}) defines a path from $s=c_{i}$ to $t=c_{j}$.
The constraint (\ref{eq:link:disjoint-edge}) ensures that the $K+1$ paths are link-disjoint.
\revise{Next,} we use the (\ref{eq:link:corePort})  constraint to ensure that each switch has at least one port for connecting to other switches and the constraint (\ref{eq:link:allPort}) defines the $max\_size$ for each switch.
The constraint (\ref{eq:link:latency}) is the latency constraint, which means that the default path (k=0) passes through at most $l_{i,j}$ switches.
The constraint (\ref{eq:link:bandwidth}) means that the bandwidth requirements of the communication flows going through the physical link $(u,v)$ \revise{must} be less than the $BW_{max}$.
The constraints (\ref{eq:link:coreLinkRouter}) and (\ref{eq:link:routerLinkcore}) ensure that each core connects exactly one switch and the constraints (\ref{eq:link:xBinary}) and (\ref{eq:link:duvBinary}) define the binary variables.

\section{Experiment}
\label{sec:experiment}

The proposed algorithms have been implemented \revise{using} C++ on a Linux 64-bit workstation (Intel 2.0 GHz, 64 GB RAM). 
\minrev{All the ILP-based formulations are solved using Gurobi \cite{gurobi}.}
In the first set of experiments, we analyzed the hardware consumption of fault-tolerant topologies.
The second set of experiments show the effectiveness of the port sharing.
In the third set of experiments, we compared the proposed method for generating link-fault-tolerant topologies with \revise{those of previous studies}.

\subsection{Hardware Cost Analysis of Fault Tolerance} \label{subsec:expNFT}

In this experiment,
the bandwidth constraint $BW_{max}$ is set at $3000MB/s$ and the maximum number of ports on switches, $max\_size$, is set to 10.
ORION 3.0 \cite{orion3} was used for estimating the switch power and the model from  \cite{huang2015lagrangian} was used for estimating the link power.

\begin{table*}[htbp]
	\centering \caption{Comparison between \textit{Non-Fault-Tolerance} \& \textit{K-Fault-Tolerance}}
	\renewcommand{\arraystretch}{1.0}
	\addtolength{\tabcolsep}{-0.5pt}
	\begin{tabular}{|c|c|c|c|c|c|c|c|c|c|c|c|}
		\hline
		\multirow{2}{*}{\centering{$K$}} & \multirow{2}{*}{Bench.} &
		\multicolumn{3}{c|}{\centering{SwitchNum}} & \multicolumn{3}{c|}{\centering{LinkNum}} & \multicolumn{3}{c|}{\centering{Power(mW)}} & \multirow{2}{*}{Time(s)}\\
		\cline{3-11}
		& &NFT & FT &$Inc$ & NFT & FT & $Inc$ &NFT & FT & $Inc$\\
		\hline
		\multirow{5}{*}{\centering{$K=1$}} & $D\_36$   &8	&10	&25.00\%	&18	&\minrev{23}	&\minrev{27.78}\%	&256.426	&\minrev{516.979}	&\minrev{101.61}\% &\minrev{2.908} \\
		& $D\_43$    & 10	&13	&30.00\%	&26	&\minrev{32}	&\minrev{23.08}\%	&330.912	&\minrev{622.819}		&\minrev{88.21}\%	&\minrev{5.102}\\
		& $D\_50$    & 12	&14	&16.67\%	&26	&\minrev{34}	&\minrev{30.77}\%	&341.88		&\minrev{692.739}	&\minrev{102.63}\%	&\minrev{8.066}\\
		& $D\_70$    & 16	&22	&37.50\%	&48	&\minrev{71}	&\minrev{47.92}\%	&574.434	&\minrev{1128.925}	&\minrev{96.53}\%	&\minrev{34.634}\\
		\cline{2-12}
		&$Average$ &  -   &  -   & 27.29\%  &   -   &    -   &\minrev{32.39}\%   &    -    &    -   & \minrev{97.24}\% &- \\
		\cline{1-12}
		\cline{1-12}
		\multirow{5}{*}{\centering{$K=2$}} & $D\_36$   &8	&15	&87.50\%	&18	&\minrev{39}	&\minrev{116.67}\%	&256.426	&\minrev{810.358}	&\minrev{216.02}\%	&\minrev{19.632} \\
		& $D\_43$    &10	&19	&90.00\%	&26	&\minrev{49}		&\minrev{88.46}\%	&330.912	&\minrev{932.885}	&\minrev{181.91}\%	&\minrev{36.554}\\
		& $D\_50$    &12	&21	&75.00\%	&26	&\minrev{59}		&\minrev{126.92}\%	&341.880	&\minrev{1099.778}	&\minrev{221.69}\%	&\minrev{68.142}\\
		& $D\_70$    &16	&36	&125.00\%	&48	&\minrev{117}	&\minrev{143.75}\%	&574.434	&\minrev{1686.243}	&\minrev{193.55}\%	&\minrev{674.757}\\
		\cline{2-12}
		&$Average$ &  -  &  -   & 94.38\%  &   -   &    -   &\minrev{118.95}\%   &    -    &    -   & \minrev{203.29}\% &-\\
		\cline{1-12}
		\cline{1-12}
		\multirow{5}{*}{\centering{$K=3$}} & $D\_36$   &8	&19	&137.50\%	&18	&\minrev{48}	&\minrev{166.67}\%	&256.426	&\minrev{1068.731}	&\minrev{316.78}\%	&\minrev{51.054}\\
		& $D\_43$    &10	&26	&160.00\%	&26	&\minrev{69}		&\minrev{165.38}\%	&330.912	&\minrev{1252.793}	&\minrev{278.59}\%	&\minrev{162.595}\\
		& $D\_50$    &12	&30	&150.00\%	&26	&\minrev{78}		&\minrev{200.00}\%	&341.88		&\minrev{1414.200}	&\minrev{313.65}\%	&\minrev{465.502}\\
		& $D\_70$    &16	&47	&\minrev{200.00}\%	&48	&\minrev{157}	&\minrev{227.08}\%	&574.434	&\minrev{2240.320}	&\minrev{290.00}\%	&\minrev{5565.602}\\
		\cline{2-12}
		&$Average$ &  -  &  -   & \minrev{170.00}\%  &   -   &    -   & \minrev{189.78}\%   &    -    &    -   & \minrev{299.76}\% &-\\
		\hline
	\end{tabular}
	\label{tab:NonFaultToleranceAnalysisWithDifferentK}
\end{table*}

Table \ref{tab:NonFaultToleranceAnalysisWithDifferentK} shows the comparison between the non-fault-tolerant topologies and the fault-tolerant topologies with $K=1$, $K=2$ and $K=3$, respectively.
The column $SwitchNum$, $LinkNum$, and $Power$ denote the number of switches, number of links, and the power, respectively.
The column $Time$ denotes the running time of the program. Additionally, the column $NFT$ and $FT$ \revise{respectively represent the non-fault-tolerant topologies and fault-tolerant topologies,} and the column $Inc$ shows the ratios.

As $K$ increased from $1$ to $3$, the power consumption is increased by \minrev{97.24}\%, \minrev{203.29}\%, and finally to \minrev{299.76}\% compared to $NFT$ ($K=0$).
Fig.\ref{fig:PowerAnalysisWithNFT} shows that the increase in the power consumption is approximately linear with $K$ for all benchmarks, because the power mainly comes from the switches and communication traffic.
The increase in the number of both switches and links is also approximately linear with $K$.
 \begin{figure}[htbp]
\small \centering
  \includegraphics[width=8.00cm,height=4.5cm]{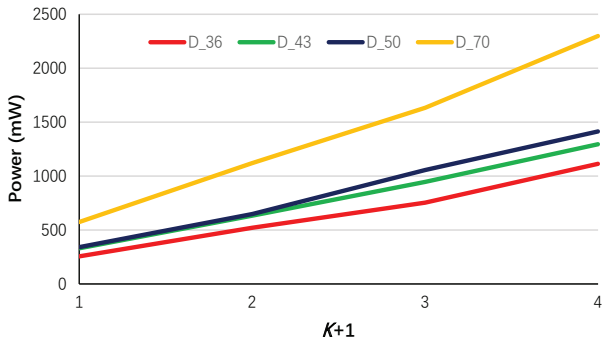}\\
  \caption{Power in different $K$-fault-tolerant topologies.}
  \label{fig:PowerAnalysisWithNFT}
\end{figure}

\begin{figure}	
	\centering
\begin{minipage}[c]{0.25\textwidth}
	\centering
	\includegraphics[height=3.5cm]{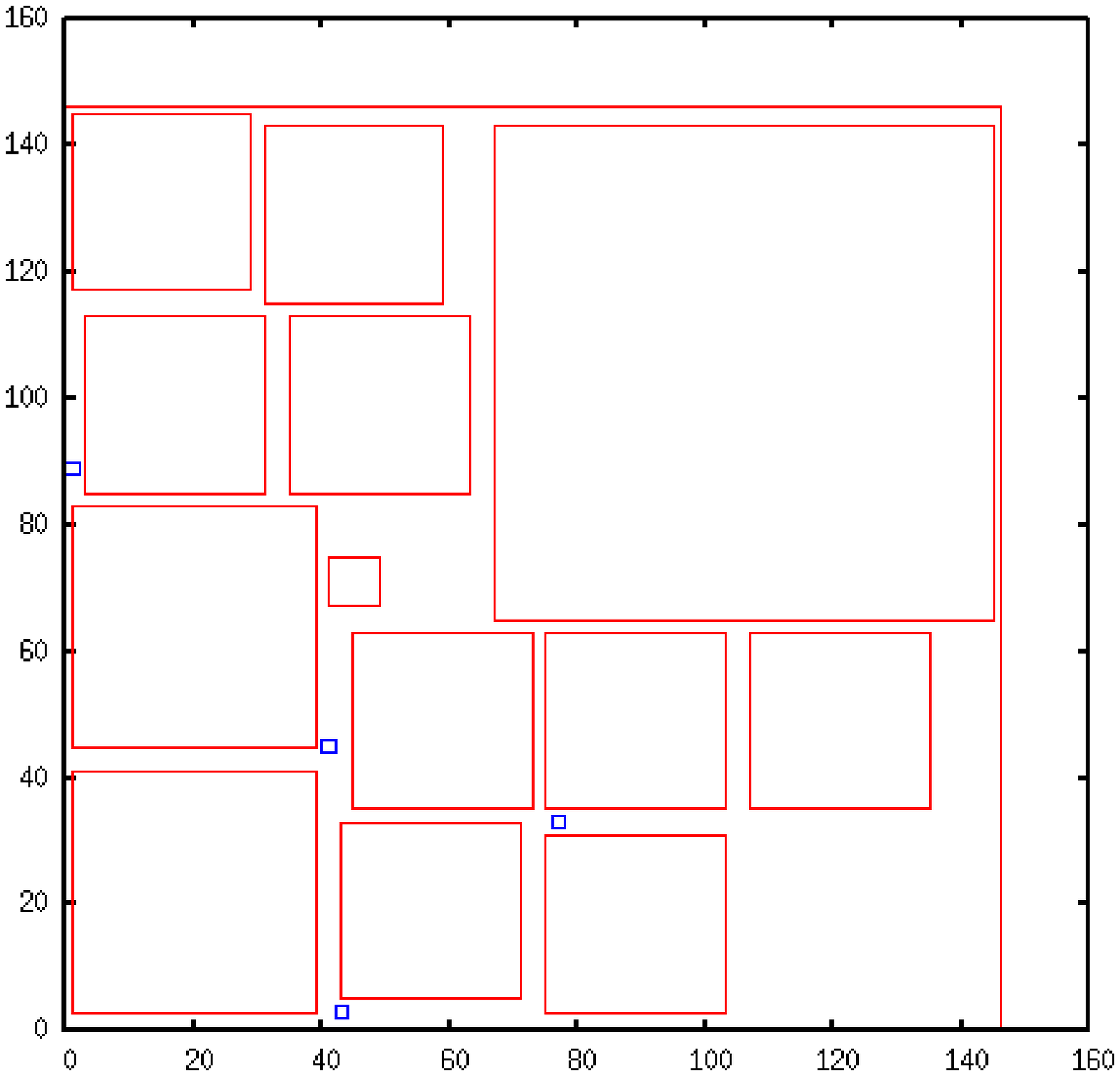}
	\subcaption{}
\end{minipage}%
\begin{minipage}[c]{0.25\textwidth}
	\centering
	\includegraphics[height=3.5cm]{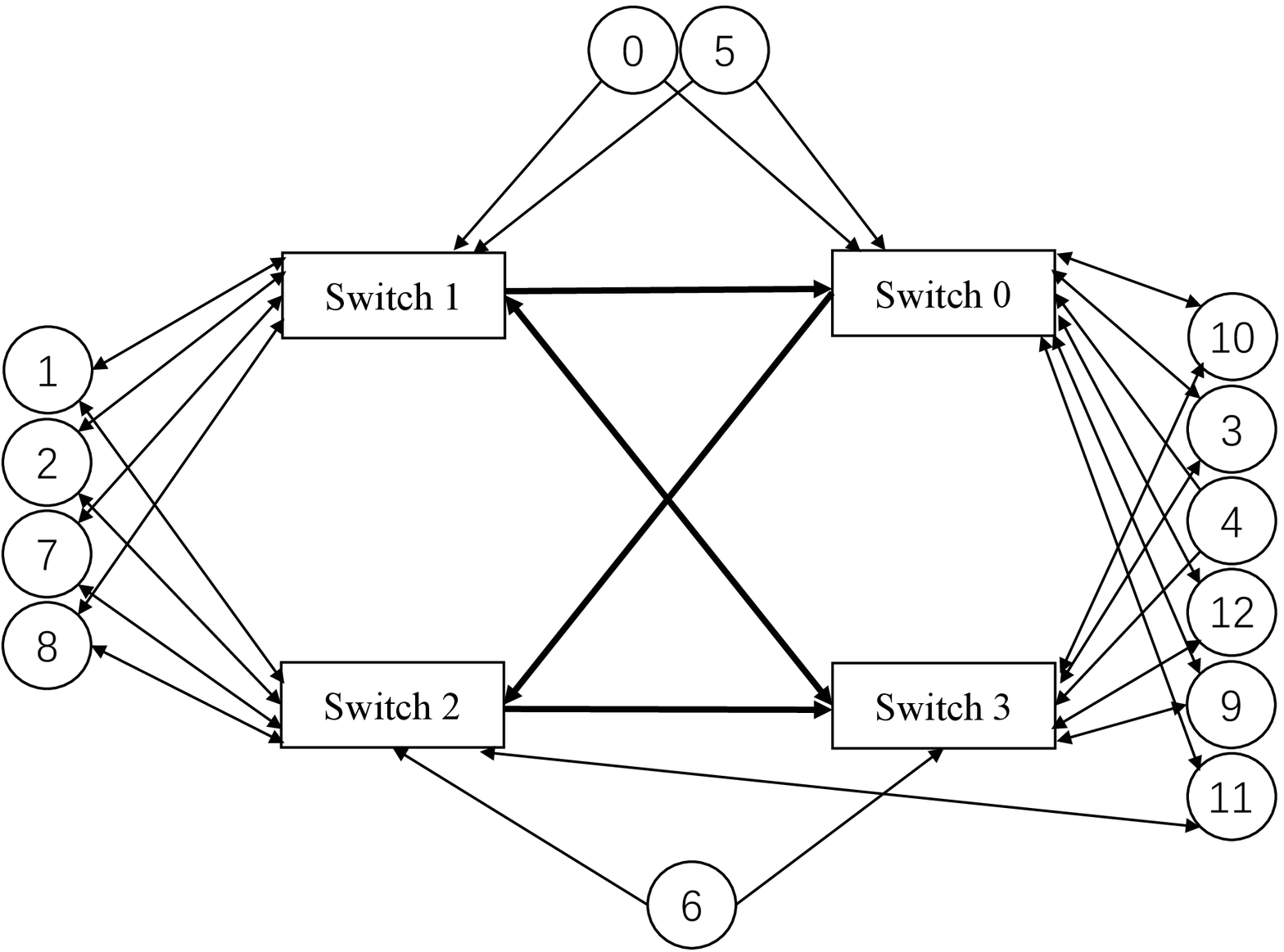}
	\subcaption{}
\end{minipage}
	\caption{One-switch-fault-tolerant topology for $MP3EncMP3Dec$: (a) floorplan. (b) topology.}\label{fig:routerFloorplan}
\end{figure}

Fig.\ref{fig:routerFloorplan} shows the result of floorplan and topology for the testbench $MP3EncMP3Dec$.

\subsection{Analysis of Port Sharing} \label{subsec:exp:ps}

Table \ref{tab:NonSharingAnalysisWithDifferentK} shows the effectiveness of the port sharing. The columns $SwitchNum$, $InportNum$, $OurportNum$, and $Power$ denote the number of switches, the number of input ports, the number of output ports, and the power, respectively. The columns $NPT$ and $PT$ represent the results with/without port sharing, and the column $Dec$ shows the reduction of $PT$ compared to $NPT$.
\begin{table*}[htbp]
\centering \caption{\small{Comparison between $Non\-Port\-Sharing$ and $Port\-Sharing$ with different $K$}}
\renewcommand{\arraystretch}{1.0}
\addtolength{\tabcolsep}{-0.5pt}
\begin{tabular}{|c|c|c|c|c|c|c|c|c|c|c|c|}
\hline
\multirow{2}{*}{\centering{$K$}} & \multirow{2}{*}{Bench.} & \multirow{2}{*}{SwitchNum} &
\multicolumn{3}{c|}{\centering{Input port}} & \multicolumn{3}{c|}{\centering{Output port}} & \multicolumn{3}{c|}{\centering{Power(mW)}} \\
\cline{4-12}
& & &NPT & PT &$Dec$ & NPT & PT &$Dec$ &NPT & PT &$Dec$\\
 \hline
\multirow{5}{*}{\centering{$K=1$}} & $D\_36$   & 10	&\minrev{87}	&\minrev{74}	&\minrev{14.94}\%	&\minrev{89}	&\minrev{77}	&\minrev{13.48}\%	&\minrev{516.979}	&\minrev{414.609}	&\minrev{19.80}\%\\
& $D\_43$    & 13	&\minrev{106}	&\minrev{91}		&\minrev{14.15}\%	&\minrev{114}	&\minrev{101}		&\minrev{11.40}\%	&\minrev{622.819}		&\minrev{506.410}	&\minrev{18.69}\%\\
& $D\_50$    & 14	&\minrev{118}	&\minrev{101}	&\minrev{14.41}\%	&\minrev{124}	&\minrev{108}		&\minrev{12.90}\%	&\minrev{692.739}	&\minrev{558.18}	&\minrev{19.42}\% \\
& $D\_70$    & 22	&\minrev{187}	&\minrev{167}	&\minrev{10.70}\%	&\minrev{197}	&\minrev{174}	&\minrev{11.68}\%	&\minrev{1128.925}	&\minrev{966.166}	&\minrev{14.42}\% \\
\cline{2-12}
&$Average$ &  -   &  -  &  -   & \minrev{13.55}\%  &   -   &    -   & \minrev{12.37}\%   &    -    &    -   & \minrev{18.08}\% \\
\cline{1-12}
\cline{1-12}
\multirow{5}{*}{\centering{$K=2$}} & $D\_36$   & 15	&\minrev{135}	&\minrev{107}	&\minrev{20.74}\%	&\minrev{138}	&\minrev{114}	&\minrev{17.39}\%	&\minrev{810.358}	&\minrev{583.529}	&\minrev{27.99}\% \\
& $D\_43$    & 19	&\minrev{160}	&\minrev{122}	&\minrev{23.75}\%	&\minrev{172}	&\minrev{142}	&\minrev{17.44}\%	&\minrev{932.885}	&\minrev{641.725}	&\minrev{31.21}\%\\
& $D\_50$    & 21	&\minrev{185}	&\minrev{140}	&\minrev{24.32}\%	&\minrev{194}	&\minrev{158}	&\minrev{18.56}\%	&\minrev{1099.778}	&\minrev{746.669}	&\minrev{32.11}\% \\
& $D\_70$    & 36	&\minrev{291}	&\minrev{238}	&\minrev{18.21}\%	&\minrev{306}	&\minrev{260}	&\minrev{15.03}\%	&\minrev{1686.243}	&\minrev{1277.790}	&\minrev{24.22}\% \\
\cline{2-12}
&$Average$ &  -   &  -  &  -   & \minrev{21.76}\%  &   -   &    -   & \minrev{17.11}\%   &    -    &    -   & \minrev{28.88}\% \\
\cline{1-12}
\cline{1-12}
\multirow{5}{*}{\centering{$K=3$}} & $D\_36$   & 19	&\minrev{176}	&\minrev{133}	&\minrev{24.43}\%	&\minrev{180}	&\minrev{143}	&\minrev{20.56}\%	&\minrev{1068.731}	&\minrev{717.322}	&\minrev{32.88}\% \\
& $D\_43$    & 26	&\minrev{217}	&\minrev{165}	&\minrev{23.96}\%	&\minrev{233}	&\minrev{162}	&\minrev{30.47}\%	&\minrev{1252.793}	&\minrev{844.043}	&\minrev{32.63}\%\\
& $D\_50$    & 30	&\minrev{246}	&\minrev{170}	&\minrev{30.89}\%	&\minrev{258}	&\minrev{186}	&\minrev{27.91}\%	&\minrev{1414.200}	&\minrev{844.494}	&\minrev{40.28}\% \\
& $D\_70$    & 47	&\minrev{389}	&\minrev{297}	&\minrev{23.65}\%	&\minrev{409}	&\minrev{314}	&\minrev{23.23}\%	&\minrev{2240.32}	&\minrev{1545.64}	&\minrev{31.01}\% \\
\cline{2-12}
&$Average$ &  -   &  -  &  -   & \minrev{25.73}\%  &   -   &    -   & \minrev{25.54}\%   &    -    &    -   & \minrev{34.20}\% \\
\hline
\end{tabular}
\label{tab:NonSharingAnalysisWithDifferentK}
\end{table*}

For one-fault-tolerance, the number of input ports, output ports, and \revise{the} power can be reduced by \minrev{13.55}\%, \minrev{12.37}\% and \minrev{18.08}\%, respectively. \revise{As $K$ increases, the results show more reduction in the switch power consumption, which demonstrates} the effectiveness of port sharing.

\subsection{Link-fault Tolerance}
\subsubsection{One-link-fault tolerance}
\label{subsec:expA}
In this subsection, we compare the proposed framework to \revise{the FTTG method in} \cite{FT}, which includes a one-link-fault switch topology generation followed by a simulated annealing based core mapping, and the \revise{de Bruijn Digraph (DBG) based method}\cite{deBG} for the one-link-fault tolerance case.

To compare our results to that of previous studies\cite{FT}\cite{deBG}, the link power of the communication flow $(i,j)$ on the edge $(u,v) \in E_{rp}$ are evaluated by $P_{lk}(i,j,u,v) = E_{bit}\cdot w_{i,j}\cdot D_{u,v}$. Notice that there are no extra costs introduced for opening a new physical link.

\textit{a. Comparison to the FTTG method: }
In this work, we stipulate that physical links are directed whereas the FTTG method uses bi-directional physical links. Hence, we degenerate the directed graph to an undirected graph by adding to the ILP formulation (\ref{eq:ilpLink}) the following constraint:
\begin{equation}
	\label{eq:link:FTTG}
	d_{uv}-d_{vu}=0, \forall (u,v)\in E_{rp}
\end{equation}

Additionally, the objective function in FTTG is to minimize the energy consumption of the network topology, which is estimated based on the shortest paths (\revise{called the} default path in \cite{FT}). To make a fair comparison to the FTTG method, we use the same objective function as follows.
\begin{equation}
\label{eq:minPre}
Min \sum_{(i,j)\in E_{cc}} \{R_{i,j}*E_{R_{bit}} + L_{i,j}*E_{L_{bit}}\}*w_{i,j}.
\end{equation}

In the formula, $E_{R_{bit}}$ represents the bit energy consumption of the switches and $E_{L_{bit}}$ represents the bit energy consumption of the unit length links \cite{FT}. $L_{i,j} = \sum_{(u,v)\in E_{rp}} d_{uv}$ is the number of physical links used by the shortest routing path of the communication flow $(i,j)$ and $R_{i,j} = L_{i,j}-1$ is the number of switches on the shortest routing path.

Six widely used benchmarks were used for the comparisons. The port number of switches $max\_size$ was set to \revise{four}, which is the same as that in \cite{FT}.
Because we cannot obtain the value of the bandwidth constraint in \cite{FT}, the bandwidth constraint $f_{max}$ is set to $3000MB/s$, which corresponds to a 32 bit physical link operating at 750MHz.
The proposed ILP-based method is applied to generate a one-fault-tolerant topology with the switch number \revise{varying from} $r_{min}$ to $r_{max}$.
For fair comparisons, we use the energy estimation model in \cite{FT}.
The energy consumption of the switches and links are set at $3.20$ $pJ/Kb$ and $4.78$ $pJ/Kb/nm$, and the length of the links is set to $1$ $mm$.

\begin{table}[htbp]
\centering \caption{\small{Comparisons to FTTG\cite{FT}}}
\renewcommand{\arraystretch}{1.0}
\addtolength{\tabcolsep}{-1.1pt}
\newcommand{\tabincell}[2]{\begin{tabular}{@{}#1@{}}#2\end{tabular}}
\begin{tabular}{|c|c|c|c|c|c|c|}
\hline
\multirow{2}{*}{Benchmark} &
\multicolumn{3}{c|}{\centering{ Energy(mJ) }} & \multicolumn{3}{c|}{\centering{Average Hop Count}} \\
\cline{2-7}
 & FTTG & Ours & $redu$ & FTTG & Ours &$redu$ \\
\hline
$MPEG4$  &61.10 &50.13  &17.95\% &1.23 &1.23 &0 \\
\hline
$VOPD$    &65.15 &49.55  &23.94\% &0.95 &0.87 &8.421\% \\
\hline
$MWD$     &19.14 &15.68  &18.08\% &0.58 &0.54 &6.897\% \\
\hline
$263 Dec$ &0.297 &0.285  &4.04\%  &0.93 &0.71 &23.655\% \\
\hline
$263 Enc$ &3.74  &3.76   &-0.54\% &0.75 &0.83 &-10.667\\
\hline
$MP3Enc$ &0.254 &0.254  &0       &0.76 &0.69 &9.211\% \\
\hline
\hline
$Average$ & -& -&10.58\% & -& -&6.25\% \\
\hline
\end{tabular}
\label{tab:FTTG}
\end{table}

The experimental results are \revise{listed} in Table \ref{tab:FTTG}.
The first column is the name of benchmark. Columns 2 and 3 provide the energy consumption of the fault-tolerant network topology generated \revise{using the} FTTG algorithm and the proposed ILP based method, respectively.
The energy is calculated \revise{based on} the shortest path between the two alternative paths, which is the default routing path \cite{FT}.
The benchmark $MP3Enc$ refers to the $MP3EncMP3Dec$ because of the limited space in the table.
Columns 5, 6, and 7 are the comparison of the two methods on the average hop count (AHC).
Furthermore, column $redu$ presents the reduction of the performance index compared with \revise{the FTTG algorithm}.
Compared with FTTG algorithm, the proposed method can reduce the energy consumption by 10.58\% on average.
In addition, the proposed method can also reduce the hop count by 6.25\%  on an average.
\revise{This} is because the switch topology and the core mapping strongly depends on each other and we formulate the two subproblems as a single ILP model, in which we can determine the best solution for the whole problems.

\textit{b. Comparison to the DBG based method: }
The DBG method \cite{deBG} generates a link-fault-tolerance topology without consideration of the position of the cores and switches. \revise{Hence,} we set the distances $D_{uv}$ to $1~mm$ to calculate the power consumption.

We implemented the $2$-D topology generation algorithm (DBG) in \cite{deBG} and made a comparison.
The port number of switches $max\_size$ is set to $10$.
Six \revise{widely used} benchmarks  and three synthetic benchmarks, $D\_36$, $D\_43$,and $D\_50$ are used in the experiments.

The experimental results are listed in Table \ref{tab:DBG}, where the average hop count is equal to the number of switches along a path plus one.
The proposed method can reduce the power and average hop count by 21.72\% and 9.35\%, respectively.
The synthetic benchmarks use one more switch compared to the DBG method\revise{; however,} the switch size is smaller, which results in a power reduction.
On the other hand, the number of links can be reduced by 45.46\% on an average. In the DBG method, each switch has two links for connecting to other switches\revise{. This results in} many redundant links.
Actually, there are some pairs of communication cores allocated on the same router which \revise{do not experience the} link-fault-tolerance issue.
In the proposed method, we consider only the necessary links between the switches.
\revise{As the number of switches is increased}, the diameter of the DGB graph increases even faster, which cause the average hop count \revise{to become} much larger, especially for the synthetic benchmarks $D\_36$, $D\_43$, and $D\_50$.

\begin{table*}[htbp]
\centering \caption{\small{Comparisons to DBG\cite{deBG}}}
\renewcommand{\arraystretch}{1.0}
\addtolength{\tabcolsep}{-0.5pt}
\newcommand{\tabincell}[2]{\begin{tabular}{@{}#1@{}}#2\end{tabular}}
\begin{tabular}{|c|c|c|c|c|c|c|c|c|c|c|c|c|}
\hline
\multirow{2}{*}{Benchmark} &
\multicolumn{3}{c|}{\centering{ SwitchNum(mJ) }} & \multicolumn{3}{c|}{\centering{LinkNum}} & \multicolumn{3}{c|}{\centering{Power(mW)}} & \multicolumn{3}{c|}{\centering{Average Hop Count}} \\
\cline{2-13}
 & DBG & Ours & reduction & DBG & Ours &reduction &DBG& Ours & reduction &DBG& Ours & reduction\\
\hline
$VOPD$ & 3 & 2 & 33.33\% & 6 & 3 & 50\% & 107.914 & 85.985 & 20.32\% & 2.20 & 2.13 & 3.18\%\\
\hline
$MPEG4$   &3 &3  &0        &6 &3 &50\%     &107.914	 &83.704	&22.43\%	&2.15	&2.31	&-7.44\%\\
\hline
$MWD$     &3 &3  &0        &6 &3 &50\%     &107.905	 &83.690	&22.44\%	&2.21	&2.15	&2.71\%\\
\hline
$MP3EncMP3Dec$ &3 &3  &0        &6 &3 &50\%     &113.810	 &84.590	&25.67\%	&2.29	&2.08	&9.17\%\\
\hline
$263 Dec$ &3 &3  &0        &6 &3 &50\%     &120.370	 &94.450	&21.53\%	&2.33	&2.14	&8.15\%\\
\hline
$263 Enc$ &3 &3  &0        &6 &3 &50\%     &107.900	 &78.680	&27.08\%	&2.15	&2.17	&-0.93\%\\
\hline
$D\_36$   &4 &5  &-25\%    &8 &5 &37.5\%   &294.581	 &238.200	&19.14\%	&2.77	&2.42	&12.64\%\\
\hline
$D\_43$   &5 &6  &-20\%    &10 &7 &30\%    &355.900 &288.426	&18.96\%	&2.95	&2.28	&22.71\%\\
\hline
$D\_50$   &6 &7  &-16.67\% &12 &7 &41.67\% &399.593	 &328.175	&17.87\%	&3.62	&2.39	&33.98\%\\
\hline
\hline
$Average$ & -& -&-6.85\% & -& -&45.46\% & - & - & 21.72\% & - & - & 9.35\%\\
\hline
\end{tabular}
\label{tab:DBG}
\end{table*}

\begin{figure}	
	\centering
\begin{minipage}[c]{0.23\textwidth}
	\centering
	\includegraphics[height=3.5cm]{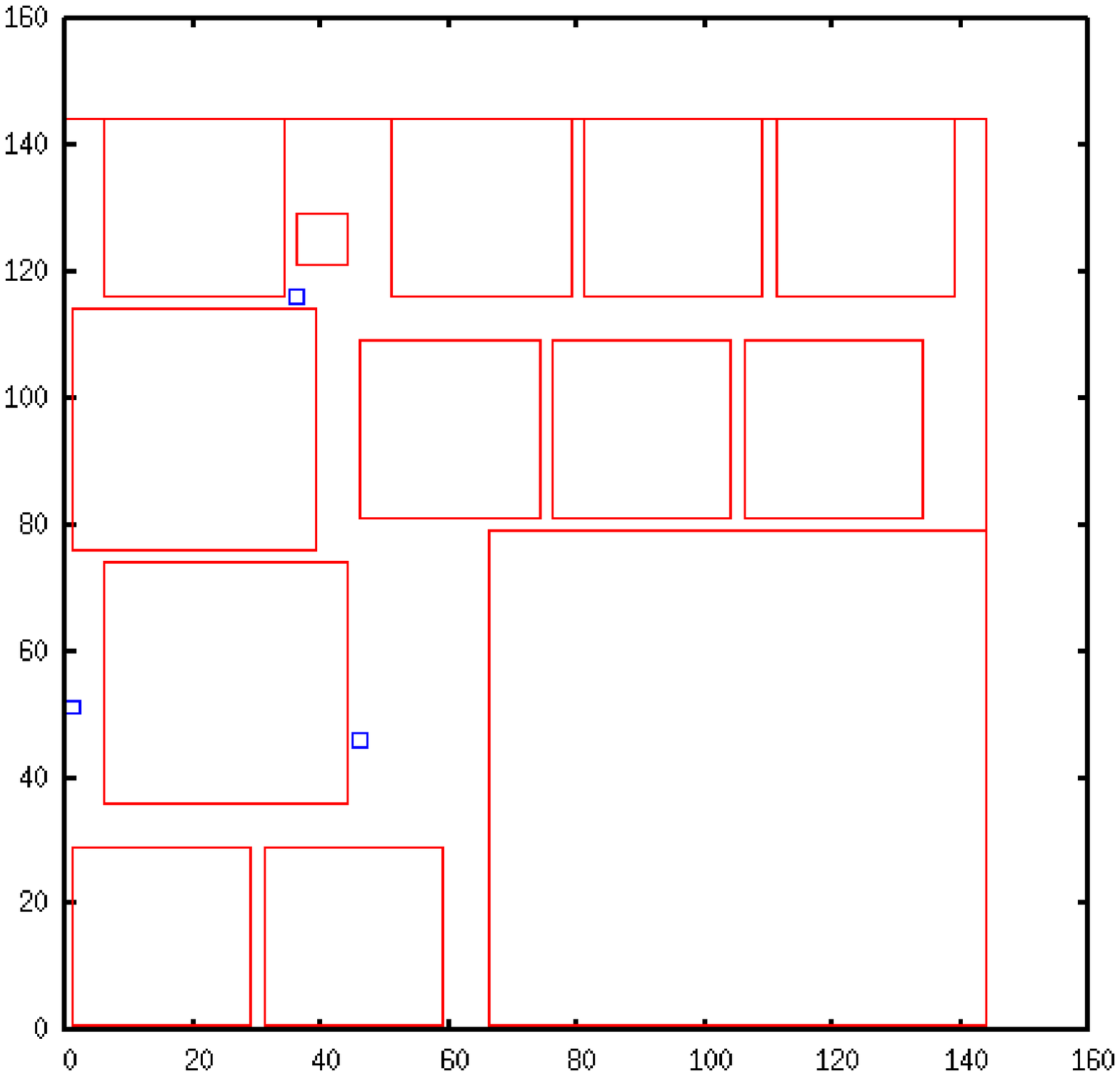}
	\subcaption{}
\end{minipage}%
\begin{minipage}[c]{0.23\textwidth}
	\centering
	\includegraphics[height=3.5cm]{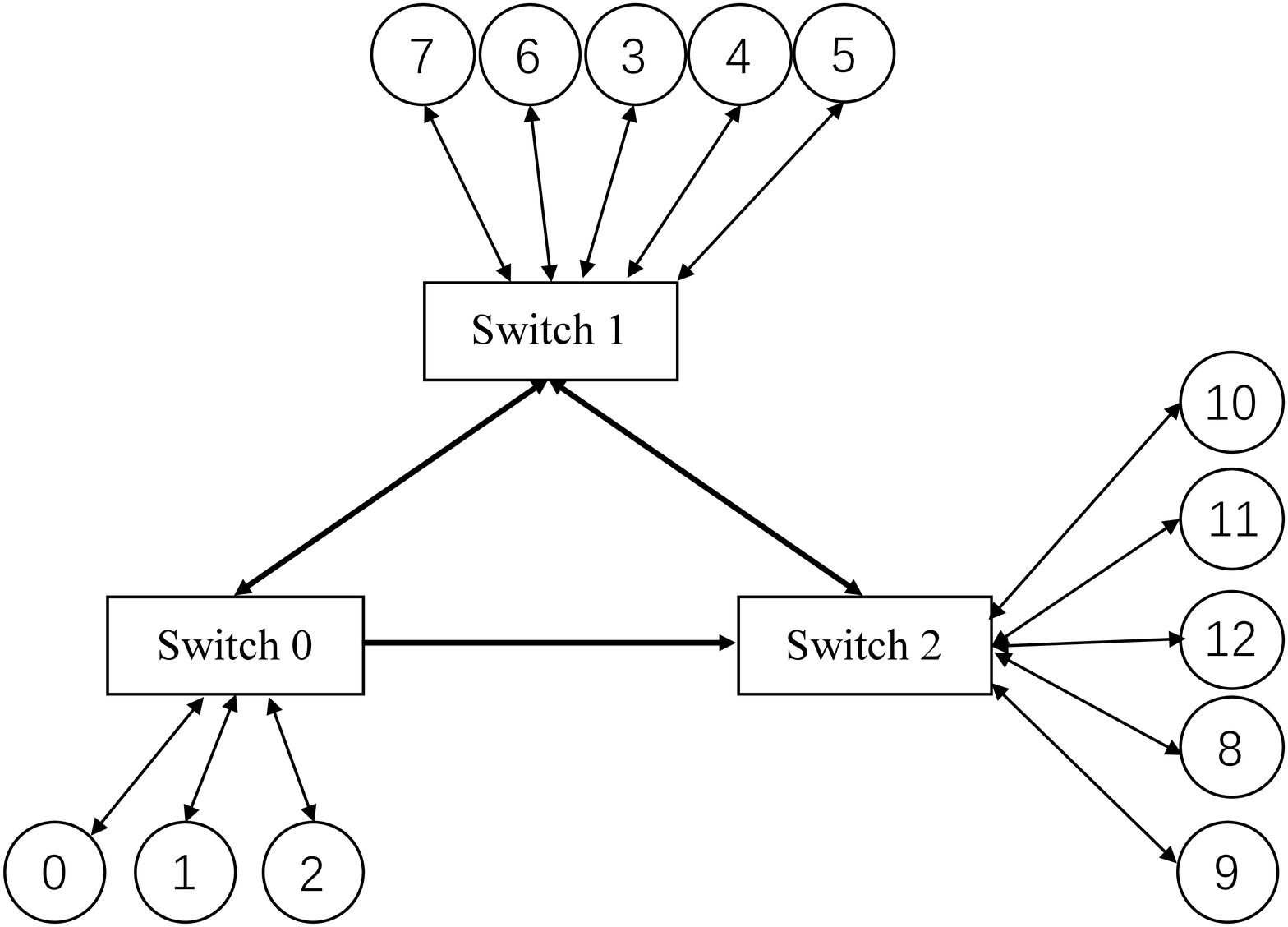}
	\subcaption{}
\end{minipage}
	\caption{One-link-fault tolerance for $MP3EncMP3Dec$: (a)Result of floorplan; (b)Result of topology}\label{fig:linkFloorplan}
\end{figure}

Fig.\ref{fig:linkFloorplan} shows the result of the floorplan and topology for the testbench $MP3EncMP3Dec$.

\subsubsection{Multiple-link-fault tolerance}
\label{subsec:exp:mft}
The proposed ILP-based method can be applied to generate $K$-link-fault tolerance network topologies.

\begin{table*}[htbp]
\centering \caption{\small{Results for $K$-link-fault ($K=0,~1,~2,~3$) tolerance}}
\renewcommand{\arraystretch}{1.0}
\addtolength{\tabcolsep}{-1.5pt}
\newcommand{\tabincell}[2]{\begin{tabular}{@{}#1@{}}#2\end{tabular}}
\begin{tabular}{|c|c|c|c|c|c|c|c|c|c|c|c|c|c|c|c|c|}
\hline
\multirow{2}{*}{Benchmark} &
\multicolumn{4}{c|}{\centering{SwitchNum}} & \multicolumn{4}{c|}{\centering{LinkNum}} & \multicolumn{4}{c|}{\centering{Power(mW)}} &\multicolumn{4}{c|}{\centering{Time(s)}}\\
\cline{2-17}
 &NFT&1FT & 2FT &3FT&NFT&1FT & 2FT &3FT &NFT&1FT & 2FT &3FT &NFT&1FT & 2FT &3FT\\
 \hline
 $MPEG4$ &3  & 3 & 4 &5 & 1 &3 &5 &7  &74.836	&78.532	&93.246	&106.269	&1.222	&10.92	&62.402	 &1091.638\\
 $VOPD$   & 3 & 3 & 4 &5 & 1 &3 &5 &7  &71.157	&79.137	&82.340	&111.381	&1.005	&7.355	&39.522	 &315.652\\
$MWD$     & 3 & 3 & 4 &5 & 1 &3 &5 &7  &67.018	&85.98	&86.067	&111.389	&1.102	&5.754	&124.97	 &1396.625\\
$263 Dec$ & 3 & 3 & 4 &5 & 1 &3 &5 &9  &83.600	&89.63	&106.25	&116.110	&1.216	&38.389	&952.437 &4314.34\\
$263 Enc$ & 3 & 3 & 4 &5 & 1 &3 &5 &7  &71.136	&82.103	&93.592	&116.370	&0.785	&3.266	&36.619	 &192.64\\
$MP3E/D$ & 3 & 3 & 4 &5 & 3 &3 &5 &9  &77.040	&82.41	&89.300	&114.930	&0.810	&6.235	&134.183 &3121.806\\
\hline
\hline
$Average$ & 1 & 1 & 1.33 & 1.67 & 1 & 2.25 & 3.75 & 5.75 & 1 & 1.119 & 1.238 & 1.521 &- &- &- &-\\
\hline
\end{tabular}
\label{tab:ourlink}
\end{table*}

Table \ref{tab:ourlink} show the results.
The benchmark $MP3E/D$ refers to the $MP3EncMP3Dec$.
The columns $SwitchNum$, $LinkNum$, and $Power$ denote the number of switches, the number of links, and the power, respectively.
The columns $NFT$, $1FT$, $2FT$, and $3FT$ denote the results of non-fault tolerance, one-fault tolerance, two-fault tolerance, and three-fault tolerance, respectively.

Compared to $NFT$, one-fault tolerance topologies use the same number of switches and three times \revise{as many} links on an average, and the power overhead of one-fault tolerance topologies is 11.9\%. Compared to $NFT$, the power consumption increases by 23.8\% in $2FT$ on average while it is 52.1\% in $3FT$, because more switches and links are used to generate fault-tolerance topologies.
The \revise{increase in} power consumption is approximately linear with $K$.

\section{Conclusions} \label{sec:conclusion}
In this paper, we presented a $K$-fault-tolerant topology generation method for ASNoC with physical link failures and switch failures.
First, an convex-cost flow and ILP based method was proposed to generate a network topology in which each communication flow has at least \revise{$K+1$ switch-disjoint routing paths, which provide $K$-fault tolerance. Second, to reduce the switch sizes, we proposed sharing the switch ports for the connections between the cores and switches, and proposed heuristic method\minrev{s} to solving the port sharing problem. Finally, we also proposed an ILP-based method to simultaneously solve the core mapping and routing path allocation problems when only the physical link failures are considered. The experimental results showed the effectiveness of the proposed method.}
%



\bibliographystyle{IEEEtran}
\bibliography{vi}

\begin{thebibliography}{10}
\providecommand{\url}[1]{#1}
\csname url@samestyle\endcsname
\providecommand{\newblock}{\relax}
\providecommand{\bibinfo}[2]{#2}
\providecommand{\BIBentrySTDinterwordspacing}{\spaceskip=0pt\relax}
\providecommand{\BIBentryALTinterwordstretchfactor}{4}
\providecommand{\BIBentryALTinterwordspacing}{\spaceskip=\fontdimen2\font plus
\BIBentryALTinterwordstretchfactor\fontdimen3\font minus
  \fontdimen4\font\relax}
\providecommand{\BIBforeignlanguage}[2]{{%
\expandafter\ifx\csname l@#1\endcsname\relax
\typeout{** WARNING: IEEEtran.bst: No hyphenation pattern has been}%
\typeout{** loaded for the language `#1'. Using the pattern for}%
\typeout{** the default language instead.}%
\else
\language=\csname l@#1\endcsname
\fi
#2}}
\providecommand{\BIBdecl}{\relax}
\BIBdecl

\bibitem{dac2007-manycore}
S.~Borkar, ``Thousand core chips: A technology perspective,'' in
  \emph{Proceedings of the 44th Annual Design Automation Conference}, 2007, pp.
  746--749.

\bibitem{intro2}
W.~J. Dally and B.~Towles, ``Route packets, not wires: On-chip interconnection
  networks,'' in \emph{Proc. 38th Annual Design Automation Conference}, 2001,
  pp. 684--689.

\bibitem{acs-2006-survey}
T.~Bjerregaard and S.~Mahadevan, ``A survey of research and practices of
  network-on-chip,'' \emph{ACM Computing Survey}, vol.~38, no.~1, June 2006.

\bibitem{tcad-2009-survey}
R.~Marculescu, U.~Y. Ogras, L.~S. Peh, N.~E. Jerger, and Y.~Hoskote,
  ``Outstanding research problems in noc design: System, microarchitecture, and
  circuit perspectives,'' \emph{IEEE Transactions on Computer-Aided Design of
  Integrated Circuits and Systems}, vol.~28, no.~1, pp. 3--21, Jan 2009.

\bibitem{truenorth2015}
F.~Akopyan, J.~Sawada, A.~Cassidy, R.~Alvarez-Icaza, J.~Arthur, P.~Merolla,
  N.~Imam, Y.~Nakamura, P.~Datta, G.~J. Nam, B.~Taba, M.~Beakes, B.~Brezzo,
  J.~B. Kuang, R.~Manohar, W.~P. Risk, B.~Jackson, and D.~S. Modha,
  ``Truenorth: Design and tool flow of a 65 mw 1 million neuron programmable
  neurosynaptic chip,'' \emph{IEEE Transactions on Computer-Aided Design of
  Integrated Circuits and Systems}, vol.~34, no.~10, pp. 1537--1557, Oct 2015.

\bibitem{neurogrid2014}
B.~V. Benjamin, P.~Gao, E.~McQuinn, S.~Choudhary, A.~R. Chandrasekaran, J.~M.
  Bussat, R.~Alvarez-Icaza, J.~V. Arthur, P.~A. Merolla, and K.~Boahen,
  ``Neurogrid: A mixed-analog-digital multichip system for large-scale neural
  simulations,'' \emph{Proceedings of the IEEE}, vol. 102, no.~5, pp. 699--716,
  May 2014.

\bibitem{neu-noc2018}
X.~Liu, W.~Wen, X.~Qian, H.~Li, and Y.~Chen, ``Neu-noc: A high-efficient
  interconnection network for accelerated neuromorphic systems,'' in \emph{23rd
  Asia and South Pacific Design Automation Conference (ASP-DAC)}, Jan 2018, pp.
  141--146.

\bibitem{intro5}
S.~Tosun and et~al., ``Application-specific topology generation algorithms for
  network-on-chip design,'' \emph{IET computers \& digital techniques}, 2012.

\bibitem{ASNoC:1}
S.~Murali, P.~Meloni, F.~Angiolini, D.~Atienza, S.~Carta, L.~Benini,
  G.~De~Micheli, and L.~Raffo, ``Designing application-specific networks on
  chips with floorplan information,'' in \emph{Proceedings of the 2006 IEEE/ACM
  International Conference on Computer-aided Design}, pp. 355--362.

\bibitem{micro2003}
C.~Constantinescu, ``Trends and challenges in vlsi circuit reliability,''
  \emph{IEEE Micro}, vol.~23, no.~4, pp. 14--19, July 2003.

\bibitem{micro2005}
S.~Borkar, ``Designing reliable systems from unreliable components: the
  challenges of transistor variability and degradation,'' \emph{IEEE Micro},
  vol.~25, no.~6, pp. 10--16, Nov 2005.

\bibitem{spectrum2011}
J.~Keane and C.~H. Kim, ``Transistor aging,'' \emph{IEEE Spectrum}, vol.~48,
  no.~5, pp. 28--33, 2011.

\bibitem{intro:redun1}
Y.~Ren, L.~Liu, S.~Yin, J.~Han, Q.~Wu, and S.~Wei, ``A fault tolerant noc
  architecture using quad-spare mesh topology and dynamic reconfiguration,''
  \emph{Journal of Systems Architecture}, vol.~59, no.~7, pp. 482--491, 2013.

\bibitem{intro:redun2}
Y.-C. Chang, C.-T. Chiu, S.-Y. Lin, and C.-K. Liu, ``On the design and analysis
  of fault tolerant noc architecture using spare routers,'' in
  \emph{Proceedings of the 16th Asia and South Pacific Design Automation
  Conference}.\hskip 1em plus 0.5em minus 0.4em\relax IEEE Press, 2011, pp.
  431--436.

\bibitem{iscas2013a}
Y.~Ren, L.~Liu, S.~Yin, Q.~Wu, S.~Wei, and J.~Han, ``A vlsi architec- ture for
  enhancing the fault tolerance of noc using quad-spare mesh topology and
  dynamic reconfiguration,'' in \emph{Proc. IEEE Int. Symp. Circuits Syst.},
  2013, pp. 1793--1796.

\bibitem{iscas2014a}
N.~Chatterjee, S.~Chattopadhyay, and K.~Manna, ``A spare router based reliable
  network-on-chip design,'' in \emph{Proc. IEEE Int. Symp. Circuits Syst.},
  2014, pp. 1957--1960.

\bibitem{intro:routingAlg}
A.~Hosseini, T.~Ragheb, and Y.~Massoud, ``A fault-aware dynamic routing
  algorithm for on-chip networks,'' in \emph{IEEE International Symposium on
  Circuits and Systems}, 2008, pp. 2653--2656.

\bibitem{toc2016}
P.~Ren, X.~Ren, S.~Sane, M.~A. Kinsy, and N.~Zheng, ``A deadlock-free and
  connectivity-guaranteed methodology for achieving fault-tolerance in on-chip
  networks,'' \emph{IEEE Transactions on Computers}, vol.~65, no.~2, pp.
  353--366, Feb 2016.

\bibitem{ASNoC:2}
K.~Srinivasan, K.~S. Chatha, and G.~Konjevod, ``Linear-programming-based
  techniques for synthesis of network-on-chip architectures,'' \emph{IEEE
  Transactions on Very Large Scale Integration (VLSI) Systems}, vol.~14, no.~4,
  pp. 407--420, 2006.

\bibitem{ASNoC:3}
S.~Yan and B.~Lin, ``Application-specific network-on-chip architecture
  synthesis based on set partitions and steiner trees,'' in \emph{Proceedings
  of the 2008 Asia and South Pacific Design Automation Conference}, pp.
  277--282.

\bibitem{ASNoC:4}
C.~Seiculescu, S.~Murali, L.~Benini, and G.~De~Micheli, ``Sunfloor 3d: A tool
  for networks on chip topology synthesis for 3-d systems on chips,''
  \emph{IEEE Transactions on Computer-Aided Design of Integrated Circuits and
  Systems}, vol.~29, no.~12, pp. 1987--2000, 2010.

\bibitem{ASNoC:5}
B.~Yu, S.~Dong, S.~Chen, and S.~Goto, ``Floorplanning and topology generation
  for application-specific network-on-chip,'' in \emph{2010 15th Asia and South
  Pacific Design Automation Conference (ASP-DAC),}, 2010, pp. 535--540.

\bibitem{ASNoC:6}
J.~Cong, Y.~Huang, and B.~Yuan, ``Atree-based topology synthesis for on-chip
  network,'' in \emph{2011 IEEE/ACM International Conference on Computer-Aided
  Design (ICCAD),}, pp. 651--658.

\bibitem{ASNoC:7}
B.~Huang, S.~Chen, W.~Zhong, and T.~Yoshimura, ``Application-specific
  network-on-chip synthesis with topology-aware floorplanning,'' in \emph{2012
  25th Symposium on Integrated Circuits and Systems Design (SBCCI)}, 2012, pp.
  1--6.

\bibitem{ASNoC:8}
W.~Zhong, S.~Chen, B.~Huang, T.~Yoshimura, and S.~Goto, ``Floorplanning and
  topology synthesis for application-specific network-on-chips,'' \emph{IEICE
  Transactions on Fundamentals of Electronics, Communications and Computer
  Sciences}, vol.~96, no.~6, pp. 1174--1184, 2013.

\bibitem{ASNoC:9}
W.~Zhong, T.~Yoshimura, B.~Yu, S.~Chen, S.~Dong, and S.~Goto, ``Cluster
  generation and network component insertion for topology synthesis of
  application-specific network-on-chips,'' \emph{IEICE transactions on
  electronics}, vol.~95, no.~4, pp. 534--545, 2012.

\bibitem{ASNoC:10}
K.~S.-M. Li, ``Cusnoc: Fast full-chip custom noc generation,'' \emph{IEEE
  Transactions on Very Large Scale Integration (VLSI) Systems}, vol.~21, no.~4,
  pp. 692--705, 2013.

\bibitem{ASNoC:11}
V.~Todorov, D.~Mueller-Gritschneder, H.~Reinig, and U.~Schlichtmann,
  ``Deterministic synthesis of hybrid application-specific network-on-chip
  topologies,'' \emph{IEEE Transactions on Computer-Aided Design of Integrated
  Circuits and Systems}, vol.~33, no.~10, pp. 1503--1516, 2014.

\bibitem{ASNoC:12}
J.~Huang, S.~Chen, W.~Zhong, W.~Zhang, S.~Diao, and F.~Lin, ``Floorplanning and
  topology synthesis for application-specific network-on-chips with
  rf-interconnect,'' \emph{ACM Transactions on Design Automation of Electronic
  Systems (TODAES)}, vol.~21, no.~3, p.~40, 2016.

\bibitem{ASNoC:13}
\BIBentryALTinterwordspacing
P.~Mukherjee, S.~D'souza, and S.~Chattopadhyay, ``Area constrained performance
  optimized asnoc synthesis with thermal‐aware white space allocation and
  redistribution,'' \emph{Integration, the VLSI Journal}, vol.~60, pp. 167 --
  189, 2018. [Online]. Available:
  \url{http://www.sciencedirect.com/science/article/pii/S0167926017301165}
\BIBentrySTDinterwordspacing

\bibitem{intro6}
A.~E. Zonouz, M.~Seyrafi, A.~Asad, M.~Soryani, M.~Fathy, and R.~Berangi, ``A
  fault tolerant noc architecture for reliability improvement and latency
  reduction,'' in \emph{IEEE 12th Euromicro Conference on Digital System
  Design, Architectures, Methods and Tools}, 2009, pp. 473--480.

\bibitem{spareRouter}
N.~Chatterjee, S.~Chattopadhyay, and K.~Manna, ``A spare router based reliable
  network-on-chip design,'' in \emph{Circuits and Systems (ISCAS), 2014 IEEE
  International Symposium on}.\hskip 1em plus 0.5em minus 0.4em\relax IEEE,
  2014, pp. 1957--1960.

\bibitem{quadSpareRouter}
Y.~Ren, L.~Liu, S.~Yin, J.~Han, Q.~Wu, and S.~Wei, ``A fault tolerant noc
  architecture using quad-spare mesh topology and dynamic reconfiguration,''
  \emph{Journal of Systems Architecture}, vol.~59, no.~7, pp. 482--491, 2013.

\bibitem{FT}
S.~Tosun, V.~B. Ajabshir, O.~Mercanoglu, and O.~Ozturk, ``Fault-tolerant
  topology generation method for application-specific network-on-chips,''
  \emph{IEEE Transactions on Computer-Aided Design of Integrated Circuits and
  Systems}, vol.~34, no.~9, pp. 1495--1508, 2015.

\bibitem{noc-stanf}
W.~J. Dally and B.~Towles, \emph{Principles and Practices of Interconnection
  Networks}.\hskip 1em plus 0.5em minus 0.4em\relax San Francisco: Morgan
  Kaufmann, 2004.

\bibitem{zhong2013floorplanning:11}
W.~Zhong, S.~Chen, and et~al., ``Floorplanning and topology synthesis for
  application-specific network-on-chips,'' \emph{IEICE Trans. Fund.}, 2013.

\bibitem{minCostFlowOfNetworkFlows}
R.~K. Ahuja, T.~L. Magnanti, and J.~B. Orlin, ``Network flows: theory,
  algorithms, and applications,'' 1993.

\bibitem{schrijver}
A.~Schrijver, \emph{{Combinatorial Optimization: Polyhedra and
  Efficiency}}.\hskip 1em plus 0.5em minus 0.4em\relax Berlin: Springer Science
  \& Business Media, 2002, vol.~24.

\bibitem{XuQi}
Q.~Xu, S.~Chen, X.~Xu, and B.~Yu, ``Clustered fault tolerance tsv planning for
  3d integrated circuits,'' \emph{IEEE Transactions on Computer-Aided Design of
  Integrated Circuits and Systems}, 2017.

\bibitem{orion3}
A.~B. Kahang, B.~Lin, and S.~Nath, ``{ORION3.0: A C}omprehensive noc router
  estimation tool,'' \emph{IEEE Embedded Systems Letters}, vol.~7, no.~2, pp.
  41--45, June 2015.

\bibitem{huang2015lagrangian}
J.~Huang, W.~Zhong, Z.~Li, and S.~Chen, ``Lagrangian relaxation-based routing
  path allocation for application-specific network-on-chips,''
  \emph{Integration, the VLSI Journal}, 2017.

\bibitem{pardalos1992branch}
P.~M. Pardalos and G.~P. Rodgers, ``A branch and bound algorithm for the
  maximum clique problem,'' \emph{Computers \& operations research}, vol.~19,
  no.~5, pp. 363--375, 1992.

\bibitem{gurobi}
\BIBentryALTinterwordspacing
I.~Gurobi~Optimization, ``Gurobi optimizer reference manual,'' 2015. [Online].
  Available: \url{http://www.gurobi.com}
\BIBentrySTDinterwordspacing

\bibitem{deBG}
K.~S.-M. Li and S.-J. Wang, ``Design methodology of fault-tolerant custom 3d
  network-on-chip,'' \emph{ACM Transactions on Design Automation of Electronic
  Systems (TODAES)}, vol.~22, no.~4, p.~63, 2017.

\end{thebibliography}

\end{document}